\documentclass[11pt, a4paper]{article}

\usepackage{stolenstyle}

\usepackage{amsmath,epsf,amssymb,latexsym,amsthm,setspace,bbm,array,pifont}

\DeclareFontFamily{U}{rsf}{}
\DeclareFontShape{U}{rsf}{m}{n}{
  <5> <6> rsfs5 <7> <8> <9> rsfs7 <10-> rsfs10}{}
\DeclareMathAlphabet\Scr{U}{rsf}{m}{n}

\def\CO#1#2{{[#1,#2]}}
\def\AC#1#2{{\{#1,#2\}}}

\def\iden{{\mathbbm 1}}

\def\rep#1{{{\boldsymbol{#1}}}}
\def\brep#1{{{\overline{\boldsymbol{#1}}}}}

\def\C{{\mathbb C}}
\def\H{{\mathbb H}}

\def\P{{\mathbb P}}
\def\R{{\mathbb R}}
\def\Z{{\mathbb Z}}

\def\Img{\operatorname{Im}}

\def\Rea{\operatorname{Re}}

\def\Tr{\operatorname{Tr}}

\def\ch{\operatorname{ch}}

\def\diag{\operatorname{diag}}

\def\rank{\operatorname{rank}}

\def\tr{\operatorname{tr}}
\def\Td{\operatorname{Td}}

\def\SO{\operatorname{SO}}
\def\SL{\operatorname{SL}}
\def\GL{\operatorname{GL}}

\def\SU{\operatorname{SU}}
\def\GU{\operatorname{U{}}}

\def\Spin{\operatorname{Spin}}
\def\GG{\operatorname{G}}
\def\GF{\operatorname{F}}
\def\GE{\operatorname{E}}

\def\so{\operatorname{\mathfrak{so}}}

\def\su{\operatorname{\mathfrak{su}}}
\def\Lu{\operatorname{\mathfrak{u}}}
\def\Le{\operatorname{\mathfrak{e}}}
\def\Lg{\operatorname{\mathfrak{g}}}
\def\LLh{\operatorname{\mathfrak{h}}}

\def\p{\partial}
\def\pb{\bar{\partial}}

\def\la{\langle}
\def\ra{\rangle}

\def\ff#1#2{{\textstyle\frac{#1}{#2}}}

\def\cA{{\cal A}}

\def\cD{{\cal D}}

\def\cF{{\cal F}}
\def\cG{{\cal G}}
\def\cH{{\cal H}}
\def\cI{{\cal I}}
\def\cJ{{\cal J}}
\def\cK{{\cal K}}
\def\cL{{\cal L}}
\def\cM{{\cal M}}

\def\cO{{\cal O}}

\def\cQ{{\cal Q}}
\def\cR{{\cal R}}
\def\cS{{\cal S}}
\def\cT{{\cal T}}

\def\cV{{\cal V}}

\def\cX{{\cal X}}

\def\ep{{\epsilon}}

\newcommand\omegah{\widehat{\omega}}

\newcommand\zetab{\overline{\zeta}}

\newcommand\lambdab{\overline{\lambda}}

\newcommand\phib{\overline{\phi}}

\newcommand\psib{\overline{\psi}}
\newcommand\omegab{\overline{\omega}}

\newcommand\omegat{\widetilde{\omega}}

\newcommand\vphi{\varphi}

\newcommand\Lambdah{\widehat{\Lambda}}

\newcommand\Psib{\overline{\Psi}}
\newcommand\Omegab{\overline{\Omega}}

\newcommand\Lambdat{\widetilde{\Lambda}}

\newcommand\Phit{\widetilde{\Phi}}

\newcommand\eh{\widehat{e}}

\newcommand\gh{\widehat{g}}

\newcommand\cb{\overline{c}}

\newcommand\hb{\overline{h}}
\newcommand\ib{\overline{\imath}}

\newcommand\vb{\overline{v}}

\newcommand\zb{\overline{z}}

\newcommand\qt{\tilde{q}}

\newcommand\Bh{\widehat{B}}

\newcommand\Eh{\widehat{E}}

\newcommand\Db{\overline{D}}

\newcommand\Kb{\overline{K}}

\newcommand\Zb{\overline{Z}}

\newcommand\Bt{\widetilde{B}}

\newcommand\Lt{\widetilde{L}}

\newcommand\Qt{\widetilde{Q}}

\def\tick{\ding{51}}
\def\cross{\ding{55}}

\def\ttick{{{\text{\tick}}}}
\def\tcross{{{\text{\cross}}}}

\def\KT{{{\text{K3}}}}

\def\nablah{{{\widehat{\nabla}}}}

\def\bpsi{{\boldsymbol{\psi}}}

\def\bca{{\boldsymbol{a}}}
\def\bcs{{\boldsymbol{s}}}

\def\bP{{\boldsymbol{P}}}
\def\bQ{{\boldsymbol{Q}}}
\def\bR{{\boldsymbol{R}}}
\def\bJ{{\boldsymbol{J}}}

\def\bomega{{\boldsymbol{\omega}}}
\def\bR{{\boldsymbol{R}}}

\def\cHh{{\widehat{\cH}}}
\def\cHt{{\widetilde{\cH}}}
\def\cAh{{{\widehat{\cA}}}}
\def\cFh{{{\widehat{\cF}}}}
\def\cRh{{{\widehat{\cR}}}}
\def\cSh{{{\widehat{\cS}}}}

\def\bJ{{\boldsymbol{J}}}

\def\Pic{\operatorname{Pic}}

\def\CS{{{\mathsf{CS}}}}

\def\Span{\operatorname{span}}

\def\bz{{{\boldsymbol{z}}}}
\def\Ppart{{{\mathsf{P}}}}

\def\sg{{{\mathsf{g}}}}
\def\bomega{{{\boldsymbol{\omega}}}}
\def\bomegat{{{\widetilde{\bomega}}}}

\def\sgt{{{\widetilde{\sg}}}}

\def\tbos{{\text{bos}}}

\def\tfree{{\text{free}}}
\def\tct{{\text{c.t.}}}

\def\tint{{\text{int}}}
\def\tfib{{\text{fib}}}
\def\tbase{{\text{base}}}

\def\ba{{\mathbf{a}}}

\def\preprint{ {{AEI-2012-048 \\ IPHT-T12/045}}}

\title{Heterotic flux backgrounds and their IIA duals}
\author[a] {Ilarion V.~Melnikov,}
\author[b] {Ruben Minasian,}
\author[a] {and Stefan Theisen}
\affiliation[a]{Max-Planck-Institut f\"ur Gravitationsphysik (Albert-Einstein-Institut)\\
 Am M\"uhlenberg 1, D-14476 Golm, Germany}
\affiliation[b]{Institut de Physique Th{\'e}orique, CEA/Saclay \\ 91191 Gif-sur-Yvette Cedex, France}
\emailAdd{ilarion@aei.mpg.de}
\emailAdd{ruben.minasian@cea.fr}
\emailAdd{theisen@aei.mpg.de}
\abstract{We study four-dimensional heterotic flux vacua 
with N=2 spacetime supersymmetry.  A worldsheet perspective is used to clarify quantization
conditions associated to the fluxes and the constraints these place on the moduli spaces of
resulting compactifications.  We propose that these vacua fit naturally in the context of
heterotic/IIA duality as heterotic duals to compactifications on K3-fibered but not elliptically fibered Calabi-Yau three-folds.
We present some examples of such potential dual pairs.}

\begin{document}

\maketitle

\section{Introduction}\label{s:intro}

String compactifications preserving N=2 super-Poincar{\'e} invariance in four dimensions
provide a demarkation line between comparatively constrained and well-understood vacua
with more supercharges and the murkier N=1 and N=0 string vacua.  In the N=2 context,
many questions that would be boring in N$>$2 theories or very difficult in N$<$2 theories
seem to be within grasp.  One of the most powerful tools at our disposal is
type II/ heterotic duality in four dimensions~\cite{Kachru:1995wm,Ferrara:1995yx}  (
standard reviews are~\cite{Aspinwall:1996mn,Aspinwall:2000fd} ).   The most familiar examples
of dual pairs are of a type IIA compactification on an elliptically fibered Calabi-Yau three-fold
and a heterotic compactification on the product manifold $T^2 \times \KT$.

The geometries involved can be constrained further by demanding that the moduli space of
the $N=2$ theory contains limiting points with local geometry that is recognizably that
of a well-behaved string compactification.  For instance, we typically assume that the moduli
space contains the weakly coupled heterotic string that is mapped to a large radius limit of
a IIA compactification on a smooth Calabi-Yau three-fold $Y$.  In this case, under relatively weak
assumptions, one can show that $Y$ must be a K3-fibered manifold~\cite{Klemm:1995tj,Aspinwall:1995vk}.
One might also wish to consider a situation where the heterotic conformal field theory is
described by a large radius non-linear sigma model.
In this case, the dual $Y$ should admit an elliptic fibration
compatible with the K3 fibration~\cite{Aspinwall:2000fd}.

What happens when the heterotic worldsheet theory does not have a large radius limit?
For instance, we might expect a generic heterotic flux compactification to have this feature;
do such theories have type II duals?  The aim of this work is to begin an exploration of these questions.
In brief, our suggestion is that perturbative heterotic flux compactifications, where the heterotic three-form flux is non-trivial at tree-level in $\alpha'$, should be naturally dual
to type IIA string theory compactified on a Calabi-Yau manifold that admits a K3 fibration but
no compatible elliptic fibration with section.  This article will mainly be concerned with the heterotic worldsheet description of N=2 vacua.
Although this subject has been explored before, we aim to give a fairly complete and comprehensible description
of various requirements for the existence of the vacuum, the geometric realization of certain
required properties of the internal superconformal theory, as well the space of marginal deformations
that preserve these properties.  

The general heterotic construction is presented in section~\ref{s:review}.  The
upshot is that the geometric structure is a principal $T^2$ bundle $X \to M$ over a K3 manifold $M$
equipped with a vector bundle $E\to X$ that admits a Hermitian Yang-Mills connection.  T-duality
suggests that the worldsheet consequences of a non-trivial $T^2$ fibration are similar to choosing
$E$ to be a line bundle over $M$.  Since this informs much of our intuition, we review the structure of
such instantons on K3 in section~\ref{s:instantons}.  

In section~\ref{s:duals} we turn to discuss potential
IIA dual descriptions of various heterotic flux vacua.  We present a few samples of interesting potential
duals, obtained by various choices of fluxes.\footnote{In earlier versions this section contained some errors; these are set right in this version.  The important qualitative modification is that it turns out to be much harder to construct examples with small $N_V$.}  We refer to these as potential duals because at this point
our evidence for duality might be fairly called ``zeroth order'' :  we construct a heterotic flux vacuum with
gauge group $G = \GU(1)^n$ and $N_H^0$ neutral hypermultiplets and
then check whether a known Calabi-Yau can realize such a massless spectrum.  In a future work we plan
to study more detailed checks of the correspondence, for instance by studying details of the vector moduli
space metric and higher derivative corrections.  

Finally, in section~\ref{s:WZW} we discuss fibered WZW models and show that the heterotic presentation of
one of the earliest models figuring in IIA/heterotic duality--- the ST model with $N_V = 2$ and $N_H = 129$~\cite{Kachru:1995wm}--- 
can be usefully thought of as a flux vacuum.  Generalizations of this construction will certainly lead to additional interesting examples of heterotic vacua.

\section{A review of heterotic N=2 compactifications} \label{s:review}
The worldsheet theory for a critical perturbative heterotic string compactification with 
a $1+3$-dimensional Minkowski vacuum decomposes into four non-interacting components:
the $(c,\cb) = (4,6)$ free (0,1) SCFT describing the Minkowski directions, a unitary
``internal'' (0,1) SCFT with $(c,\cb) = (c',9)$, a left-moving current algebra with
$(c,\cb) = (22-c',0)$, and the (0,1) $bc-\beta\gamma$ system with $(c,\cb)=(-26,-15)$.  
The complete theory should admit a heterotic GSO projection
leading to a tachyon-free spectrum and modular invariance.
This structure is further restricted in vacua with spacetime 
supersymmetry.  Vacua with N=1 spacetime supersymmetry require the internal theory
to be a (0,2) SCFT with integral R-charges~\cite{Hull:1985jv,Sen:1986mg,Banks:1987cy},
and N=2 spacetime supersymmetry, the case of interest for this paper, requires the right-moving
superconformal algebra (SCA) to decompose into a product of a $\cb = 3$ and $\cb = 6$ algebras with,
respectively, (0,2) and (0,4) supersymmetry~\cite{Banks:1988yz,Lauer:1988aw}. 

The spacetime gauge symmetry provides an important and relatively straightforward characterization of any perturbative heterotic vacuum.\footnote{Additional, non-perturbative sources of gauge symmetry certainly
exist~\cite{Witten:1995gx} and have important implications for, among other things, type II/heterotic duality~\cite{Aspinwall:1997ye}.}  There are two ways to construct vertex operators for the emission of spacetime
gauge bosons.  If we label the Minkowski (0,1) multiplets as $(\vec{X}, \vec{\chi})$, where $\vec{\chi}$ are 
the four right-moving fermions, and denote the spin field for the $\beta$-$\gamma$ system by $e^{-\vphi}$, 
then we have, in the $-1$-picture~\cite{Friedan:1985ge,Polchinski:1998rr},
\begin{align} \vec{\cV}_{\text{g.b.}} = e^{-\vphi} \bJ_{\!L} \vec{\chi} e^{i \vec{k} \cdot \vec{X}} \quad \text{or} \quad \vec{\cV}'_{\text{g.b.}} =  e^{-\vphi} \p \vec{X} \Psi_{\!R} e^{i \vec{k} \cdot \vec{X}} ,
\end{align}
where $\bJ_{\!L}$ is a left-moving current (belonging either to the internal theory or the additional left-moving current algebra) with conformal weights $(h,\hb) = (1,0)$, and $\Psi_{\!R}$ is a right-moving fermion with
$(h,\hb) = (0,1/2)$.  The latter operator is the lowest component of a (0,1) superconformal current 
algebra (SCCA).  The existence of SCCAs leads to strong constraints on the theory~\cite{Dixon:1987yp}.  For instance,
a theory with a non-abelian SCCA does not have any massless fermions in the spectrum, while an abelian SCCA
is equivalent to a free compact (0,1) SCFT, and its presence implies that the compactification has a non-chiral 
spectrum; moreover, every massless fermion must be neutral with respect to an abelian SCCA.

A unitary N=2 SCA with c=3 and integral R-charges has a canonical decomposition into two abelian N=1 SCCAs.  This follows from
a Sugawara decomposition of the generators $J, G^\pm, T$ into a pair of free fermions $\Psi, \Psib$ and bosonic currents $\p Z, \p \Zb$:
\begin{align}
\label{eq:N2} 
J = \Psi\Psib, \quad G^+ = i\sqrt{2} \Psi \p \Zb, \quad G^- = i\sqrt{2} \p Z, \quad
T = - \p Z \p \Zb - \ff{1}{2} ( \Psi \p \Psib + \Psib \p \Psi).
\end{align}
As a consequence of this, we immediately see that the massless spectrum of a perturbative heterotic vacuum with N=2 spacetime supersymmetry has two canonical gauge bosons associated to the two SCCAs.  All massless
fermions, including the gravitini, are neutral with respect to these, and furthermore, these symmetries cannot
be either spontaneously broken or enhanced to a non-abelian symmetry within perturbation theory.  Of course this is not
surprising from the spacetime point of view, where we also expect two canonical gauge bosons --- the graviphoton and the partner of the heterotic axio-dilaton.  The former belongs to the gravity multiplet, while the latter is in a vector multiplet.\footnote{The natural multiplet structure for the axio-dilaton is the ``vector--tensor'' multiplet~\cite{deWit:1995zg}; it can be dualized to a standard vector multiplet, at least as far as perturbation theory is concerned.}  Note that in what follows, \emph {when we speak of ``the gauge symmetry'' of an N=2 theory, we will leave out the graviphoton.} 

Having described some general features of perturbative N=2 compactifications, we will
now illustrate how they arise in the case that the internal SCFT can be described by a
heterotic non-linear sigma model.  As we will not restrict ourselves to weakly coupled 
NLSMs, we should note that our discussion will be a bit formal; for the cases at
hand, we assume that at least some basic properties of the SCFT are accurately reflected
by the fields and Lagrangian of the NLSM --- namely, the existence of certain chiral symmetries,
and the central charges can be read off from the fields and Lagrangian.  As our examples will have a large
amount of worldsheet supersymmetry, our assumptions are not unreasonable and
perhaps even testable by carefully studying and constraining the structure of quantum corrections to the
worldsheet theory.

\subsection{The (0,1) heterotic non-linear sigma model}
The classical theory is easily presented in (0,1) superspace.\footnote{Our worldsheet and superspace conventions are those of~\cite{Polchinski:1998rr}.}  We work on a genus zero Euclidean worldsheet $\Sigma$ with canonical bundle $K_\Sigma$ and denote the superspace coordinates by $\bz \equiv (z;\zb,\theta)$.  The superspace covariant derivatives are
\begin{align}
\cD & \equiv \p_\theta + \theta \pb, \qquad \cQ \equiv \p_\theta -\theta\pb, \nonumber\\ 
\cD^2 &= \pb, \qquad \cQ^2 = -\pb, \qquad \AC{\cD}{\cQ} = 0. 
\end{align}
Supersymmetry transformations with parameter $\xi$ act as
\begin{align}
\delta_\xi \bz = \delta_\xi (z,\zb,\theta) \equiv  (\xi \cQ z,\xi \cQ \zb,\xi \cQ\theta) = (0, -\xi\theta, \xi),
\end{align}
and the (0,1) supercharge $\bQ_1$ acts on a superfield $X$ by
\begin{align}
\delta_\xi X = \xi\bQ_1 \cdot X  \equiv - \xi \cQ X.
\end{align} 
We will have use for two types of multiplets: 
\begin{align}
\Phi^\mu = \phi^\mu + i\theta\psi^\mu\quad\text{(bosonic)}, \qquad
\Lambda^A = \lambda^A + \theta L^A\quad\text{(fermionic)}.
\end{align}
As usual, $\phi^\mu(z,\zb)$, $\mu=1,\ldots,6$, are local coordinates for the map from $\Sigma$ to the target space $X$, while their partners $\psi^\mu$ are sections of $\Kb_{\Sigma}^{1/2} \otimes \phi^\ast(T_X)$.  The $\lambda^A$, $A = 1,\ldots, 32$, are the left-moving fermions and the $L^A$ are auxiliary fields; $\lambda \equiv (\lambda^1,\ldots,\lambda^{32})^T$ is valued in $K_{\Sigma}^{1/2} \otimes \phi^\ast (E)$, where $E \to X$ is a vector bundle with structure group $G_E \subset \SO(32)$ or $G_E \subset \SO(16)\times \SO(16)$.

The classical action is specified in terms of metric $g$, B-field $B$ on $X$, and a connection $\cA$ on $E$.  
We will focus exclusively on connections $\cA$ that have a regular embedding in $\so(32)$ 
or $\so(16)\times\so(16)$, so that we can think of $\cA$ as valued in the  appropriate fundamental
representation.  More general cases require a more sophisticated worldsheet 
treatment~\cite{Distler:2007av}.   The superspace action is then (we set $\alpha' = 2$)
\begin{align}
\label{eq:eucahsuper}
S = \frac{1}{4\pi} \int d^2 z d\theta \left\{ (g_{\mu\nu}+B_{\mu\nu}) \p \Phi^\mu \cD\Phi^\nu 
- \Lambda^T ( \cD \Lambda + \cA_\mu \cD\Phi^\mu \Lambda) \right\},
\end{align}
and the equations of motion are 
\begin{align}
\label{eq:ceom}
\cD \Lambda & = - \cA_\mu \cD\Phi^\mu \Lambda, \nonumber\\
g_{\nu\rho} \p\cD\Phi^\rho & = - (\Gamma_{\nu\lambda\mu}-\ff{1}{2} d B_{\nu\lambda\mu}) \p\Phi^\lambda\cD\Phi^\mu +\ff{1}{2} \Lambda^T \cF_{\nu\mu} \Lambda \cD\Phi^\mu. 
\end{align}
The component action, with auxiliary fields $L$ eliminated by their equations of motion, is
\begin{align}
S &= \frac{1}{4\pi} \int d^2z \left\{ (g_{\mu\nu} +B_{\mu\nu}) \p \phi^\mu \pb\phi^\nu + 
g_{\mu\nu} \psi^\mu\p\psi^\nu + \p \phi^\lambda \psi^\mu\psi^\nu (\Gamma_{\mu\lambda\nu}-\ff{1}{2} dB_{\mu\lambda\nu})  \right. \nonumber\\
&\Bigl.\qquad\qquad\qquad +\lambda^T ( \pb \lambda + \pb \phi^\mu\cA_\mu \lambda)
- \ff{1}{2} \lambda^T \cF_{\mu\nu}\lambda \psi^\mu\psi^\nu 
\Bigr\},
\end{align}
where $\cF = d\cA + \cA^2$ is the curvature of the connection $\cA$.
Note that while the kinetic terms for the left- and right-moving fermions appear to have a very different form,
we can use a vielbein $e^a_\mu$ and its inverse $E^{a\mu}$ to express the action in terms of frame bundle fermions
$\bpsi^a \equiv e^a_\mu \psi^\mu$ with the result
\begin{align}
g_{\mu\nu} \psi^\mu\p\psi^\nu + \p \phi^\lambda \psi^\mu\psi^\nu (\Gamma_{\mu\lambda\nu}-\ff{1}{2} dB_{\mu\lambda\nu})  = \bpsi^T ( \p \bpsi + \p \phi^\mu\cS^{-}_\mu \bpsi), 
\end{align}
where $\cS^\pm$ denote the spin connection $\omega$ twisted by $H = dB$:
\begin{align}
\label{eq:Spm}
\cS^{\pm ab}_\lambda = \omega^{ab}_\lambda \pm \ff{1}{2} E^{a\sigma} E^{b\nu} H_{\sigma\lambda\nu}.
\end{align}

\subsection{The Green-Schwarz mechanism and the one-loop effective action}
The classical action is invariant under gauge transformations 
\begin{align}\delta_\ep \lambda = \ep \lambda \qquad \text{and} \qquad
\delta_\ep \cA = - \nabla \ep = - d\ep - \CO{\cA}{\ep},
\end{align} 
where the gauge parameter $\ep$ is pulled back from the target space.  Similarly, the
action is invariant under Lorentz transformations\footnote{Note that  $\nabla$ denotes
both the gauge and Lorentz-covariant derivative in the target space.} 
\begin{align}\delta_\kappa \bpsi = \kappa \bpsi \qquad \text{and} \qquad
\delta_\kappa \omega = -\nabla \kappa = - d\kappa - \CO{\omega}{\kappa}.
\end{align} 
As is well-known, these transformations are in general anomalous~\cite{Hull:1985jv,Hull:1986xn}.
Demanding that the symmetries are preserved requires non-trivial transformations of the $B$-field,
and the resulting Bianchi identity leads to the global constraint $p_1(T_X) = p_1(E)$.  This is of
course the worldsheet manifestation of the Green-Schwarz mechanism.

Even if the Bianchi identity is satisfied, we might worry whether the counter-terms required to
preserve the gauge invariance are (0,1) supersymmetric.  Fortunately, this is the case~\cite{Hull:1986xn},
with the result a delicate combination of local counter-terms and non-local non-covariant terms in the
effective action.  We will have use for the particular form of these terms, so we review the details of the
computation of~\cite{Hull:1986xn} in appendix~\ref{app:background}.  
The result of the background field computation is that to quadratic order in $\cA$ and $\cS^+$ the 
non-covariant contribution from the one-loop effective action is a sum of three terms: 
\begin{align} \Delta S = \Delta S_{\cA} + \Delta S_{\cS^+}  - S_{\tct}.
\end{align}
$S_{\tct}$ is a local term
\begin{align}
S_{\tct} = -\frac{1}{8\pi} \int d^2 z d\theta \left[ \tr\{\cA_\mu\cA_\nu\} - \tr\{\cS^+_\mu \cS^+_\nu\} \right] \p\Phi^\mu \cD \Phi^\nu.
\end{align}
Note that $\tr\{\cdots\}$ denotes either the fundamental of $\so(32)$ or $\so(6)$, depending on whether
the argument is a gauge or Lorentz object.  As the name suggests, this contribution is canceled by adding
$S_{\tct}$, a finite local counter-term, to the action.  The ``truly non-local'' contributions are
\begin{align}
\label{eq:Snloc}
\Delta S_{\cA}  &=- \int  \frac{ d^2z_1 d^2 z_2}{(4\pi)^2 z_{12} } d\theta_2 d\theta_1 
\tr\{\cA_{1\mu} d\cA_{2\lambda\rho}\} \cD_1 \Phi_1^\mu \cD_2 \Phi_2^\lambda \p_2\Phi_2^\rho, \nonumber\\ 
\Delta S_{\cS^+}  &= +\int  \frac{ d^2z_1 d^2 z_2}{(4\pi)^2 z_{12} } d\theta_2 d\theta_1 
\tr\{\cS^+_{1\mu} d\cS^+_{2\lambda\rho}\} \cD_1 \Phi_1^\mu \cD_2 \Phi_2^\lambda \p_2\Phi_2^\rho.
\end{align}
Here the subscripts denote the superspace coordinates of the fields and derivatives; for example,
$\cA_{1\mu} \equiv \cA_{\mu} (\Phi(\bz_1))$, $\cD_1 \equiv \p_{\theta_1} + \theta_1 \pb_1$, etc.
Note the obvious but useful fact that $\Delta S_{\cS^+}$ is obtained from $\Delta S_{\cA}$ by
switching the overall sign and replacing $\cA \to \cS^+$.

While the effective action is explicitly (0,1) supersymmetric, it is not gauge-invariant.  The supersymmetry
identity
\begin{align}\cD_1 z_{12}^{-1} = 2\pi (\theta_1-\theta_2) \delta^2(z_{12},\zb_{12})\end{align}
shows that under linearized transformations $\delta_\ep \cA = - d\ep$ and $\delta_\kappa \cS^+ = -d\kappa$,
the action transforms by a local term
\begin{align}
\label{eq:gaugevar}
\delta\Delta S =\frac{1}{8\pi} \int d^2 z d\theta (\tr\{\ep d\cA_{\mu\nu} \} -\tr\{\kappa d\cS^+_{\mu\nu}\}) \p\Phi^\mu \cD\Phi^\nu. 
\end{align}
This variation is canceled by postulating the B-field transformation
\begin{align}
\label{eq:delB}
\delta B = - \frac{1}{2} \tr\{\ep d\cA\} + \frac{1}{2} \tr\{\kappa d\cS^+\}.
\end{align}
That means the gauge-invariant three-form is
\begin{align}
\cH \equiv dB - \frac{1}{2} \CS_3(\cA) + \frac{1}{2}\CS_3(\cS^+), \qquad 
\CS_3(\cA) \equiv \tr\{ \cA d\cA+\ff{2}{3} \cA^3 \}.
\end{align}
The result has been obtained to quadratic order in $\cA$ and $\cS^+$, but we expect (and will assume)
that inclusion of the higher order terms will lead to the non-linear covariant form. 

\subsection{Anomalies and relevant characteristic classes}
 Having reviewed the (0,1) NLSM and the mechanism of anomaly cancelation, we will now
 discuss some global conditions necessary for consistent perturbative heterotic 
 compactifications in the RNS formalism.
 
 Restoring $\alpha'$ and evaluating $d\cH$ leads to the familiar form of the Bianchi identity
\begin{align}
\label{eq:Bianchi}
d \cH = \frac{\alpha'}{4} (\tr\{\cR_+^2\}-\tr\{\cF^2\} ),
\end{align}
where $\cR_+ = d\cS^+ + \cS_+^2$ is the curvature of the twisted spin connection.  This leads to
a topological condition on the first Pontryagin classes of $E$ and $T_X$.  As the normalization of
these will play a role in our analysis, we will quickly review a few basic facts about these classes.
This is standard and classic, see e.g.~\cite{Atiyah:1978wi,Eguchi:1980jx} for differential aspects and~\cite{MR0440554} for the algebraic topology.

Given a connection $\cA$ for a principal $G$-bundle $P \to X$, the first Pontryagin class is a basic
topological invariant constructed from the curvature $\cF = d\cA + \cA^2$:
\begin{align}
p_1 (\Lg) = -\frac{1}{8\pi^2 h_{\Lg}} \Tr\{ \cF^2 \} \in H^4(X,\Z).
\end{align}
Here $\Lg$ is the Lie algebra of $G$, $h_{\Lg}$ is the dual Coxeter number, and $\Tr\{\cdots\}$, the trace in the adjoint representation, is
normalized so that the highest root has length-squared $2$.  

In this work we are interested in heterotic gauge bundles that are constructible by starting with
a free fermion representation of $\GE_8\times\GE_8$ or $\Spin(32)/\Z_2$ and gauging a subset of
the global symmetries.  Thus it is natural to think of a rank $k$ vector
bundle $E$ with associated principal bundle as above, and we will write $p_1(E)$ for the corresponding
Pontryagin class.  The Bianchi identity~(\ref{eq:Bianchi}) implies $p_1(E) = p_1(T_X)$ in $H^4(X,\R)$.

In general, a compactification that solves the Bianchi identity still suffers from a global anomaly~\cite{Witten:1985ga,Freed:1986zx,Distler:1986wm} if a Stiefel-Whitney class $w_1(E)$ or
$w_2(E)$ is non-zero.  For a Hermitian bundle $E$ this anomaly is absent provided 
\begin{align}c_1(E) = 0 \mod 2.\end{align} 
The spacetime origin of this condition is not too hard to understand.  Consider, for example, a compactification of the $\GE_8\times \GE_8$ string with bundle $E$ and $\Lg_E\subset \so(16)\subset\Le_8$.  The 
ten-dimensional $\Le_8$ gauge bosons decompose as $\rep{248} = \rep{120} \oplus \rep{128}$ under the $\so(16)$, and all of these correspond to (possibly massive) states in the theory; however, in order for an $\so(16)$ bundle to have spinor representations, $E$ must have vanishing second Stiefel-Whitney class --- $w_2(E) = 0$~\cite{Lawson:1989sp}.\footnote{We also require $w_1(E) = 0$; however, that is a much weaker condition:  for instance, it is satisfied for any compact simply connected base space, or whenever $E$ is Hermitian.}  If $E$ is Hermitian, then $w_2(E) = c_1(E) \mod 2$, and we recover the familiar condition on the first Chern class.
For even more mundane reasons $X$ must be spin, so that $w_1(T_X) = w_2(T_X)  = 0$ as well.  Finally, note that for an orientable vector bundle $E$ we have~\cite{MR0440554}
\begin{align} p_1(E) = w_2(E)^2 \mod 2.\end{align}
Consequently, if $w_1(E) = w_2(E) = 0$, then $p_1(E) \in H^4(X,2\Z).$
The Bianchi identity is then required to hold in integral cohomology~\cite{Witten:1985ga,Freed:1986zx} as
\begin{align}
\frac{1}{2} p_1(E) - \frac{1}{2} p_1(T_X) = 0 \in H^4(X,\Z).
\end{align}

\subsection{Constraints from (0,2)+(0,4) supersymmetry} \label{ss:0204}
We will now review the conditions under which (0,1) supersymmetry of the NLSM is enhanced to the full (0,2)+(0,4) necessary for N=2 spacetime supersymmetry.\footnote{The connection between (0,2) supersymmetry enhancement in the NLSM and N=1 spacetime supersymmetry was explored much earlier in~\cite{Hull:1985jv,Sen:1986mg}.}  These were considered in~\cite{Melnikov:2010pq},
but the presentation we will now give will be a bit simpler and will close a small gap in the arguments of~\cite{Melnikov:2010pq}.  

A good starting point for the constraints is to demand that the NLSM give a realization of the $\cb = 3$
algebra of~(\ref{eq:N2}).  In order for this symmetry to be manifest in the geometric description, the
metric $g_{\mu\nu}$ must have two commuting isometries $\p/\p\theta^I$, which means the target space $X$
takes the form of a $T^2$ fibration $X \to M$, with metric
\begin{align}
g =\gh_{ij}(y) dy^i dy^j + \cG_{IJ}(y) \Theta^I \Theta^J, \qquad \Theta^I \equiv d\theta^I + A^I_i(y) dy^i,
\end{align}
where the $y^i$ are local coordinates on $M$, the connections $A^I$ describe the fibration structure, 
and $\cG_{IJ}$ is some (possibly base-dependent) metric in the fiber directions.  Similarly, the
gauge connection and B-field can be decomposed as 
\begin{align}
\cA &= \cAh+ \bca_I \Theta^I = \cAh_i(y) dy^i +\bca_I(y)\Theta^I, \nonumber\\
B &= \Bh + \Bt_I \Theta^I + \ff{1}{2} b\ep_{IJ} \Theta^I \Theta^J =
\ff{1}{2} \Bh_{ij}(y) dy^i dy^j + \Bt_{Ii}(y) dy^i \Theta^I + \ff{1}{2} b\ep_{IJ} \Theta^I \Theta^J.
\end{align}
The tree-level superspace action~(\ref{eq:eucahsuper}) splits as $S = S_\tbase+S_\tfib$ with\footnote{In this section we will omit the superspace measure $d^2z~d\theta$ when it is not likely to cause confusion.} 
\begin{align}
4\pi S_\tbase & = \int  \left[ (\gh_{ij} + \Bh_{ij} )\p\Phi^i \cD\Phi^j - \Lambda^T (\cD\Lambda + \cAh_i \cD\Phi^i \Lambda) \right], \nonumber\\
4\pi S_\tfib & = \int  \left[ (\cG_{IJ}+b\ep_{IJ}) D_z \Phi^I \cD_\theta \Phi^J 
+ \Bt_{Ij} ( \p\Phi^j \cD_{\theta} \Phi^I-D_z\Phi^I \cD\Phi^j ) + \Lambda^T \bca_I \Lambda \cD_\theta \Phi^I \right],\end{align}
where $\Phi^i$ ($\Phi^I$) correspond to the base (fiber) coordinates, and the covariant derivatives are
\begin{align}
D_z\Phi^I \equiv \p\Phi^I + A^I_i(\Phi)\p\Phi^i, \quad
\Db_{\zb}\Phi^I \equiv \pb\Phi^I + A^I_i(\Phi)\pb\Phi^i, \quad
\cD_\theta \Phi^I &\equiv \cD \Phi^I + A^I_i(\Phi)\cD\Phi^i.
\end{align}
Expanding these in components we find
\begin{align}
D_z\Phi^I &= D_z\phi^I + i \theta (\p \Psi^I + F^I_{ij} \psi^i \p\phi^j) ,\nonumber\\
\Db_{\zb}\Phi^I &= \Db_{\zb}\phi^I + i \theta (\pb \Psi^I + F^I_{ij} \psi^i \pb\phi^j) ,\nonumber\\
\cD_\theta \Phi^I &= i\Psi^I + \theta (\Db_{\zb} \phi^I -\ff{1}{2} F^I_{ij} \psi^i\psi^j),
\end{align}
where $F^I = dA^I$, $\Psi^I \equiv \psi^I + A^I_i \psi^i$, and the bosonic derivatives
are $D_z \phi^I = \p\phi^I + A^i_i \p\phi^i$ and similarly for $\Db_{\zb}\phi^I$.  Note that all of
these quantities are invariant under the Kaluza-Klein gauge symmetries 
$\delta_f \Phi^I = f^I(\Phi^i)$ and $\delta_f A^I = -d f^I.$

We can give a similar expansion of the non-local terms in~(\ref{eq:Snloc}).  We have
\begin{align}
\label{eq:DSA}
\Delta S_{\cA} &= -\int d^2z_2 d\theta_2 \int d^2z_1 d\theta_1 \frac{1}{(4\pi)^2 z_{12}} \tr\{ X_{\cA 1} Y_{\cA 2} \}, \quad \text{where} \nonumber\\[0.2cm]
X_{\cA} & \equiv \cAh_i \cD\Phi^i + \bca_I \cD_\theta \Phi^I, \nonumber\\
Y_{\cA} & \equiv (d\cAh_{ij} +\bca_I F^I_{ij} )\cD\Phi^i \p\Phi^j 
+ \bca_{I,j} (\cD\Phi^j D_z \Phi^I - \cD_{\theta} \Phi^I \p \Phi^j).
\end{align}
To obtain $\Delta S_{\cS^+}$ from $\Delta S_{\cA}$ write $\cS^+ = \cSh^+ + \bcs^+_I \Theta^I$; now flip the sign of $\Delta S_{\cA}$ and substitute $\cAh \to \cSh^+$, $\bca \to \bcs^+$. 

\subsubsection*{The torus symmetries}
The chiral symmetries necessary for the $\cb = 3$ algebra require that the background
be chosen such that $\p\Psi^I = 0$ up to equations of motion and that
\begin{align}
\label{eq:torus}
\delta_v \Phi^I = v^I(\zb), \qquad \delta_v \Lambda = -v^I(\zb) \bca_I \Lambda
\end{align} 
are symmetries of the action.  Under a variation $\delta\Phi^I$ we find $\delta S_\tbase = 0$, and
\begin{align}
\label{eq:DSfib}
4\pi \delta S_\tfib & = \int \delta\Phi^I 
\left[ -2 \cG_{IJ} \p\cD_\theta \Phi^J + ( d\Bt_{Ijk}+(\cG_{IJ}-b\ep_{IJ})F^J_{jk}) \p\Phi^j\cD\Phi^k 
\right. \nonumber\\
& \qquad \qquad \qquad \left.
-(\cG_{IJ}+b\ep_{IJ})_{,k} \p\Phi^k \cD_\theta \Phi^J 
-(\cG_{IJ}-b\ep_{IJ})_{,k} \cD\Phi^k D_z \Phi^J
- \cD (\Lambda^T \bca_I \Lambda)  \right].
\end{align}
We also find
\begin{align}
\delta \Delta S_{\cA} & = \frac{1}{8\pi} \int \delta\Phi^I \tr\{\bca_{I} Y_{\cA} + \p\bca_{I} X_{\cA}\} 
+ \int_2\int_1 \frac{1}{(4\pi)^2 z_{12}} \delta \Phi_1^I \tr\{ \cD_1 \bca_{1I} (Y_{2\cA} - \p_2 X_{2\cA}) \},
\end{align}
as well as a similar term for $\delta \Delta S_{\cS^+}$.

To obtain $\p\Psi^I = 0$ as an equation of motion requires the variation of the full action
to be proportional to $\delta\Phi^I E_{IJ} \p\cD_\theta \Phi^J$ for some invertible $E_{IJ}$.  Clearly
this places strong constraints on the background geometry.  To start, consider the contributions 
to~(\ref{eq:DSfib}) that involve the $\Lambda$ multiplets.  Using the $\Lambda$ equations of motion
these can be rewritten as
\begin{align} -\cD(\Lambda^T \bca_I \Lambda) = -\Lambda^T \nablah_i \bca_I \Lambda \cD\Phi^i +
\Lambda^T \CO{\bca_I}{\bca_J} \Lambda \cD_\theta \Phi^J .\end{align}
Here $\nablah = d + \cAh$ is the gauge-covariant derivative on the base.
These contributions cannot be canceled by any others, so we obtain our first constraints on the
background:  
\begin{align}
\label{eq:acond}
\nablah \bca_I = 0, \qquad \CO{\bca_I}{\bca_J} = 0.
\end{align}
These conditions imply that $\cF$ has no fiber components:
\begin{align} \cF = \cFh + \bca_I F^I.\end{align}


Next we will examine the non-local terms in the variation.  Here we face an awkward issue since
the terms quadratic in $\cA$ and $\cS^+$ are not by themselves explicitly covariant.  On the other 
hand, we expect the conditions on the background to be covariant, so we will assume that inclusion
of the higher order contributions will yield covariant expressions.  With this assumption we see
that since $\nablah \bca_I = d\bca_I + \CO{\cAh}{\bca_I}$, we can neglect derivatives 
of $\bca_I$ in $\delta \Delta S_{\cA}$.  The variation of $\Delta S_{\cA}$ is then purely local:
\begin{align}
\label{eq:dDSA}
\delta \Delta S_{\cA} & = - \frac{1}{8\pi} \int \delta\Phi^I \tr\{\bca_{I} (d\cAh_{jk} +\bca_{J} F^J_{jk})\} \cD\Phi^k \p\Phi^j .
\end{align}
The remaining non-local term from $\delta \Delta S_{\cS^+}$ must vanish by itself, which leads to 
$d \bcs^+_I = 0$ to leading order in the background.  The obvious covariant form of this condition is
$\nablah \bcs^+_I = 0$, and the remaining variation of $\delta \Delta S_{\cS^+}$ is
%
\begin{align}
\delta \Delta S_{\cS^+} & =  \frac{1}{8\pi} \int \delta\Phi^I \tr\{\bcs^+_{I} (d\cSh^+_{jk} +\bcs^+_{J} F^J_{jk})\} \cD\Phi^k \p\Phi^j .
\end{align}
Since now all terms in $\delta \Delta S$ are proportional to $\p\Phi^i\cD\Phi^j$, the terms proportional
to $\p\Phi^j \cD_\theta\Phi^K$ and $\cD\Phi^j D_z\Phi^K$ in~(\ref{eq:DSfib}) must vanish by themselves.  Thus, we find another
constraint:
\begin{align}
\label{eq:Gb}
\cG \quad\text{and}\quad b \quad\text{are constant over $M$.}
\end{align}
The latter condition means
\begin{align} dB = d\Bh -\Bt_I F^I + (d\Bt_I -b \ep_{IJ} F^J) \Theta^I,\end{align}
and expanding the gauge-invariant three form $\cH$ in a similar horizontal-vertical decomposition
$\cH = \cHh + \cHt_I \Theta^I$ we find
\begin{align}
\label{eq:bigH}
\cHh & = d \Bh - \Bt_I F^I -\ff{1}{2} (\CS_3(\cAh) +\tr\{\bca_I \cAh\} F^I ) 
+\ff{1}{2} (\CS_3(\cSh^+) +\tr\{\bcs^+_I \cSh^+\} F^I ), \nonumber\\
\cHt_I & = d\Bt_I -b\ep_{IJ}F^J 
-\ff{1}{2} (\tr\{\bca_I (2\cFh +\bca_J F^J)\} - d\tr\{\bca_I \cAh\} ) \nonumber\\
&\qquad\qquad\qquad\qquad
+\ff{1}{2}( \tr\{\bcs^+_I (2\cRh^+ +\bcs^+_J F^J) \} - d\tr\{\bcs^+_I \cSh^+\}).
\end{align}
Comparing the remaining terms in the variation with $\cHt_I$, we see that
\begin{align}
4\pi \delta S & = \int \delta \Phi^I \left[ -2 \cG_{IJ} \p\cD_\theta\Phi^J 
+(\cG_{IJ} F^J_{jk} + \cHt_{Ijk}) \p\Phi^j\cD\Phi^k \right].
\end{align}
Thus, we will obtain the desired equation of motion $\p \Psi^I = 0$ if
\begin{align}
\label{eq:Hvert}
\cHt_{I} = - \cG_{IJ} F^J.
\end{align}
The conditions in~(\ref{eq:acond}, \ref{eq:Gb}, \ref{eq:Hvert}), together with $\nablah \bcs^+_I = 0$,
are also sufficient to ensure that the action possesses the expected chiral symmetry~(\ref{eq:torus}).

Using~(\ref{eq:Hvert}) and~(\ref{eq:Gb}) we find another important simplification on the 
background: $\bcs^+_I = 0$.  To see this, write the metric $g$ and (torsion-free, metric-compatible) 
spin connection $\omega$ with base(fiber) frame indices $a,b$ ($A,B$) as
\begin{align}
 g = \eh^a \otimes \eh^a + \cG_{IJ} \Theta^I \otimes \Theta^J, \qquad
 \omega = \omegah + \omegat_I \Theta^I.
\end{align}
A short computation shows $\omegah^{a}_{~b}$ is the spin connection for the base metric $\gh$,
and the remaining non-vanishing components of $\omegah$, $\omegat$ are
\begin{align}\omegah^A_{~b} = -\ff{1}{2} \eh^a F^A_{ab} , \qquad
\omegah^b_{~A} = +\ff{1}{2} \eh^a F^B_{ab} \cG_{BA}, \qquad
\omegat^a_{I b} = \ff{1}{2} F^{A}_{ba} \cG_{AI}.\end{align}
Plugging this into the expression for $\cS^+$ in~(\ref{eq:Spm}) yields
\begin{align}
\bcs^{+ab}_I = \ff{1}{2} F^A_{ba} \cG_{AI} +\ff{1}{2} d\Bt_{Iba}.
\end{align}
We expect the proper covariant form $\bcs^+_I$ to be given by replacing $d\Bt_{Iba} \to \cHt_{Iba}$,
and from~(\ref{eq:Hvert}) we conclude that 
\begin{align}\bcs^+_I = 0.
\end{align}  
This means that the curvature $\cR_+$ has no fiber components, and since the same is true of $\cF$, the characteristic classes in the Bianchi identity are purely horizontal:
\begin{align} d\cH = -\ff{1}{2} \tr\{ (\cFh+\bca_I F^I)^2\} +\ff{1}{2} \tr\{\cR_+^2\}.\end{align}

\subsubsection*{Remaining conditions}
We will now discuss the remaining conditions that lead to the NLSM with a manifest
(0,2)+(0,4) symmetry~\cite{Melnikov:2010pq}.  Having ensured that the fiber fermions
$\Psi^I$ behave as the free fermions of the (0,2) algebra, the $\GU(1)_R$ symmetry of the (0,2)
algebra is generated by 
\begin{align}r \cdot \Psi^I = -i \cI^I_J \Psi^J,\qquad r\cdot \psi^i = 0.\end{align}
For $r$ to be a symmetry of the action $\cI$ must be constant and $\cG$-compatible.

The $\SU(2)_R$ symmetry
generators $R_a$ leave the $\Psi^I$ invariant and act on the base fermions by
\begin{align} R_a \cdot \psi^i = -i \cK^i_{aj} \psi^j - i \widetilde{\cK}^i_{a J} \psi^J.\end{align}
Requiring that the action is invariant leads to $\widetilde{\cK}_{a} = 0$ as well as
\begin{align}
\label{eq:Kcond}
 &\cK^i_{ak} \gh_{ij} + \cK^i_{aj} \gh_{ik} = 0, \qquad
 \cK^i_{ak} F^J_{ij} + F^J_{ki} \cK^i_{aj} = 0, \qquad
 \cK^i_{ak} \cFh_{ij} +  \cFh_{ki} \cK^i_{aj}, \nonumber\\
&\nablah^+_j \cK^i_{ak} \equiv \nablah_j \cK^i_{ak} + \ff{1}{2} (\cHh_{j~m}^i \cK^m_{ak} - \cHh_{j~k}^m \cK^i_{am})  = 0.
\end{align}
Here $\cFh = d\cAh + \cAh^2$.  In order to realize the $\SU(2)$ algebra on the fields we should
also have $\CO{\cK_a}{\cK_b} = 2 \ep_{abc} \cK_c$.

Recall the manner in which the (0,1) supersymmetry is enhanced to (0,2)~\cite{Hull:1985jv,Sen:1986mg}.
Given the R-symmetry generator $\bR$, the known supercharge $\bQ_1$, and the translation generator
$\bP = \bQ_1^2 = \pb$, we can define a second supersymmetry generator 
$\bQ_2 \equiv i\CO{\bQ_1}{\bR}$ and demand that these operators close to the (0,2) algebra
with non-trivial commutators
\begin{align}
\CO{\bR}{\bQ_A} = i\ep_{AB} \bQ_B, \qquad \AC{\bQ_A}{\bQ_B} = 2 \delta_{AB} \bP.
\end{align}
It is not hard to show using the Jacobi identity that this will hold if $\bR$ and $\bP$ commute and
 $\bQ_1 = i\CO{\bR}{\bQ_2}$.  
 
 In the case at hand there are a number of (0,2) sub-algebras with $\bR = \pm r +R_a$; closure requires $\cI$ and $\cK_a$ to be complex structures for the fiber and base directions,
respectively.  In a similar fashion we can construct the remaining generators of (0,2)+(0,4)
and check closure of the full algebra.  This does not lead to additional
constraints~\cite{Melnikov:2010pq}.  Since we will perform a similar computation in
section~\ref{s:WZW}, we will not discuss it further here.

\subsubsection*{Geometric interpretation}
Using $\cK_a^2 = -\iden$ and $\CO{\cK_a}{\cK_b} = 2 \ep_{abc} \cK_c$, we find
$\cK_a \cK_b = -\delta_{ab} \iden + \ep_{abc} \cK_c.$  This, together with the
metric compatibility condition, shows that the base manifold $M$ is a hyper-Hermitian 
surface~\cite{Joyce:2007rh} with a triplet of Hermitian forms $(J_a)_{ij} = \cK^k_{ai}\gh_{kj}$.
These can be shown to satisfy 
\begin{align}d J_a = \beta \wedge J_a,\end{align}
 where $\beta$ is a closed $1$-form
determined solely by the base metric $\gh$.  The remaining conditions in~(\ref{eq:Kcond}) constrain 
$\cHh = -\ast_{\gh} \beta$ and the $F^J$ and $\cFh$ to be (1,1) with respect to all three
complex structures.  The latter is equivalent to $F^J$ and $\cFh$ being anti-self-dual.

Compact hyper-Hermitian surfaces were classified in~\cite{Boyer:1988hm}.\footnote{We are interested in compact backgrounds; there has also been recent work on related non-compact heterotic backgrounds, e.g.~\cite{Becker:2008rc,Carlevaro:2011mn}.}  The result is that
$M$ is conformal to one of the following: $T^4$ with its flat metric, $\KT$ with its hyper-K\"ahler metric, or a Hopf surface.
Examination of the Bianchi identity shows that $M = T^4$ requires the fibration to be 
trivial~\cite{Becker:2006et}.  Hopf surfaces~\cite{Kodaira:1981cx} are excluded for a more
subtle reason:  the resulting total space $X$ does not admit a conformally balanced metric, 
or equivalently, does not have a holomorphically trivial canonical 
bundle~\cite{Melnikov:2010pq}.\footnote{From the spacetime point of view triviality of the canonical
bundle is a consequence of the vanishing dilatino variation necessary for N=1 spacetime supersymmetry~\cite{Strominger:1986uh}; 
it also emerges as a condition of (0,2) superconformal invariance~\cite{Hull:1986kz,Nibbelink:2012wb}.}   

So, to summarize, (0,2)+(0,4) supersymmetry implies that the NLSM  target space $X$ is
either $T^6$ without flux, or it is a (possibly trivial) principal $T^2$ bundle over $M=\KT$ with ASD connections
$A^I$.  The gauge bundle data is an ASD connection $\cAh$ together with a choice of covariantly constant and commuting ``Wilson lines'' $\bca_I$.  Duality arguments~\cite{Dasgupta:1999ss}, as well as explicit 
existence results~\cite{Fu:2006vj,Becker:2006et} show that the requisite connections and metric $\gh$ exist.
The resulting NLSM describes a heterotic vacuum with N=2 supersymmetry at one loop in $\alpha'$.

\subsection{Moduli and flux quantization} \label{ss:fluxquantization}
Given the existence of a perturbative N=2 vacuum, the next natural question is the characterization
of its vector- and hypermultiplet moduli spaces.  While describing the full geometry is not so simple,
at least finding the dimensions is reasonably straightforward.  To orient the discussion in the flux
case, consider the trivial fibration $X = T^2 \times \KT$.  In this case the moduli are arranged as follows.
\begin{enumerate}
\item The gauge-neutral hypermultiplets correspond to moduli of the ASD connection $\cAh$ and
the geometric (including the $B$-field) moduli of the K3. 
\item The axio-dilaton resides in a privileged vector multiplet; we described how the corresponding 
gauge boson arises from the right-moving SCCA.
\item The remaining vector moduli consist of the constant Wilson lines $\bca_I$ in the Cartan subalgebra of the spacetime gauge group, as well as the two parameters $\tau$ and $\rho$ for the complex structure and complexified K\"ahler form on $T^2$.
\end{enumerate}

How does this picture change in a flux vacuum?  The axio-dilaton structure remains unchanged. The gauge-neutral hypermultiplets correspond to moduli of $\cAh$ and the base geometry that preserve the (0,2)+(0,4) conditions.  The resulting restrictions on the geometric moduli are well-understood:  they are essentially the same as those that arise in the case of abelian instantons discussed in section~\ref{ss:abinst}.  In this section we will concentrate on the vector moduli associated to the torus.
  
These are clearly modified since the left-moving
symmetries $\delta \phi^I = v^I(z)$ are explicitly broken by the non-trivial curvatures $F^I$.\footnote{This assumes that the $F^I$ are linearly independent; a left-moving symmetry and corresponding gauge boson can be preserved if the $F^I$ are linearly dependent in $H^2(M,2\pi \Z)$.}  On the other hand, nothing in our construction so far has placed any restrictions on the torus metric and B-field $\cG$ and $b$.  As we will now argue, the requisite restrictions arise due to quantization conditions on $\cH$.   In general such quantization conditions arise from a proper interpretation of the heterotic $B$-field~\cite{Witten:1999eg}, and the case at hand is a nice illustration of the general notions.  For us the basic point is that unlike in the familiar type II case, where $B$ is a connection on an abelian gerbe, so that $\cH \in  H^3(X,4\pi^2\alpha'\Z)$~\cite{Alvarez:1984es,Rohm:1985jv}, in the heterotic case $B$ is a torsor over the group of connections on abelian gerbes:  i.e. given a $B$ for fixed $E$ and $X$, any other $B'$ for the same data arises as $B' = B + B_{\text{g}}$ for some unique gerbe connection $B_{\text{g}}$.\footnote{A precise formulation of this may be found in~\cite{Hopkins:2002rd}; we thank S.~Katz for pointing out this reference.}

\subsubsection*{Significance of $\cHt_I = -\cG_{IJ} F^J$}
To describe the quantization conditions, we first return to~(\ref{eq:bigH}) and rewrite it by using
$\cHt_I = -\cG_{IJ} F^J$ and $\bcs^+_I = 0$.  Restoring $\alpha'$, this leads to
\begin{align}
d \left(\Bt_I+\ff{\alpha'}{4} \tr\{\bca_I \cAh\} \right) & = -(\cG^\ast_{IJ}-b\ep_{IJ} ) F^J +\ff{\alpha'}{2} \tr\{\bca_I\cFh \}, \nonumber\\
\cHh & = d\Bh - \left(\Bt_I + \ff{\alpha'}{4} \tr\{\bca_I \cAh\} \right)F^I 
-\ff{\alpha'}{4} \CS_3(\cAh) + \ff{\alpha'
}{4} \CS_3(\cSh^+),
\end{align}
where 
\begin{align}
\cG_{IJ}^\ast \equiv \cG_{IJ} -\ff{\alpha'}{4} \tr\{\bca_I\bca_J\}.
\end{align}
Note that $\nablah \bca_I = 0$ implies $\cG_{IJ}^\ast$ is constant and $\tr\{\bca_I \cFh\}$ is closed.

Let us consider the gauge, Lorentz, and gerbe transformations of $B$ in more detail.  Parametrizing the transformations by, respectively, $\ep$, $\kappa$, and the one-form $\Lambda = \Lambdah +\Lambdat_I \Theta^I$, we find that the components of $B$ transform as
\begin{align}
\delta \Bh & = d\Lambdah +(\Lambdat_I -\ff{\alpha'}{4} \tr\{\ep \bca_I\})F^I-\ff{\alpha'}{4} \tr\{\ep d\cAh\} +\ff{\alpha'}{4} \tr\{\kappa d\cSh^+\}, \nonumber\\
\delta \Bt_I & = d\Lambdat_I -\ff{\alpha'}{4} \tr\{\ep d\bca_I\}, \qquad \delta b = 0.
\end{align}
We can usefully untangle some of these transformations via the redefinitions
\begin{align}
\Bt'_I \equiv \Bt_I + \ff{\alpha'}{4} \tr\{\bca_I \cAh\}, \qquad
\Lambdat'_I \equiv \Lambdat_I -\ff{\alpha'}{4}\tr\{\ep \bca_I\},
\end{align}
which lead to
\begin{align}
\delta \Bh  = d\Lambdah +\Lambdat'_I F^I-\ff{\alpha'}{4} \tr\{ \ep d\cAh\} +\ff{\alpha'}{4} \tr\{\kappa d\cSh^+\}, \qquad \delta b = 0.
\end{align}
while the $\Bt'_I$ satisfy
\begin{align}
d\Bt'_I = -(\cG^\ast_{IJ}-b\ep_{IJ} ) F^J + \ff{\alpha'}{2} \tr\{\bca_I\cFh \}, \qquad \delta \Bt'_I = d\Lambda'_I.
\end{align}
Evidently, $\Bt'_I$ behave as connections on two line bundles, and the curvatures $d\Bt'_I$ have to be separately quantized.  To determine the precise quantization conditions, we note that the Bianchi identity takes the form
\begin{align}
d \cHh = - d\Bt'_I F^I -\ff{\alpha'}{4} \tr\{\cFh^2\} +\ff{\alpha'}{4} \tr\{\cRh_+^2\} .
\end{align}
The cohomological Bianchi identity is then given by
\begin{align}
\label{eq:Bianchibase}
-\frac{d \Bt'_I }{2\pi\alpha'} ~ \frac{F^I}{2\pi} 
 + \ff{1}{2} p_1(\Eh) -\ff{1}{2} p_1(T_M) = 0 \in H^4(M,\Z).
\end{align}
Since the last two terms are  quantized, the first term must be quantized as well, and we see that
the integrality is preserved under shifts of $d\Bt'_I$ by elements of $H^2(M, 2\pi\alpha' \Z)$.  Thus,
we conclude that the appropriate quantization condition for $d\Bt'_I$ is
\begin{align}
\label{eq:finalfluxquantization}
d\Bt'_I = -(\cG^\ast_{IJ}-b\ep_{IJ} ) F^J + \ff{\alpha'}{2} \tr\{\bca_I\cFh \}  \in H^2(M,2\pi\alpha' \Z).
\end{align}
Setting for the moment $\tr\{\bca_I \cFh\} = 0$, we see that for linearly independent $F^I$ this leads to a quantization of $\cG^\ast$ and $b$.

It is straightforward to include the modifications when $\bca_I \cFh \neq 0$; however, giving a general discussion of the possibilities is a bit awkward.  Instead of doing so, we will point out two important cases.  First, when $\cAh$ is an irreducible connection, i.e. where the holonomy of the connection is the expected group $G_E$, then $\nablah \bca_I = 0$ requires $\bca_I$ to be constant and valued in $\Lg'$, the commutant of $\Lg_{E}$ in $\so(16)\times\so(16)$ or $\so(32)$.  It is easy to see this when $G_E = \SO(k)$ or $G_E = \GU(k)$; to illustrate this, we will examine the former case.  Decomposing the connection and Wilson lines as
\begin{align}
\cA = \begin{pmatrix} \cAh & 0 \\ 0 & 0 \end{pmatrix}, \qquad\qquad
\bca_I = \begin{pmatrix} a_I & b_I \\ -b^T_I & a'_I \end{pmatrix} ,
\end{align}
we find that $\nablah \bca_I = 0$ holds iff $da'_I =0$, while $a_I$ and $b_I$ are covariantly constant and, in particular, invariant under parallel transport.  If either of these is non-zero, then it must be that the holonomy group of the connection is a proper subgroup of $\SO(k)$, and hence the connection is reducible.  

Thus, we see that when the connection is irreducible $\bca_I \cFh = 0$.
 In this case the spacetime gauge algebra is $\Lg'$, and the commuting constant Wilson lines $a'_I \in \Lg'$ parametrize the Coulomb branch for the $\Lg'$ vector multiplets.   Note, however, that the quantization condition does involve these $\bca_I$, since it is $\cG^\ast$ and not $\cG$ that is quantized.

When the connection $\cAh$ is reducible, $\tr\{\bca_I \cFh\}$ need not be zero.  A simple example of this is obtained by taking $G_{E} = \SO(2)$. 
In the fundamental representation appropriate to the free fermion construction we have (ignoring the commutant)
\begin{align} \cFh = \begin{pmatrix} 0 & \cF \\ -\cF & 0 \end{pmatrix},
\qquad
\bca_I =  \begin{pmatrix} 0 & w_I \\ -w_I & 0 \end{pmatrix},
\end{align}
so that the quantization condition reads
\begin{align}
 -(\cG^\ast_{IJ}-b\ep_{IJ} ) F^J - \alpha' w_I \cF \in H^2(M,2\pi\alpha'\Z).
\end{align}
As long as $F^I$ and $\cF_1$ define linearly independent classes there are separate quantization conditions on $\cG^\ast$, $b$ and $w_I$.  However, if there is a linear dependence, say $\cF = m_J F^J$, then the quantization conditions are weaker:
\begin{align} -\cG_{IJ}^\ast + b\ep_{IJ} -2 \alpha' w_I m_J \in \alpha'\Z ,\end{align}
leaving the $w_I$ unfixed.  However, in this case we also expect an additional massless gauge boson, and the $w_I$ will be the scalars in the corresponding vector multiplet.

\section{Instantons on K3} \label{s:instantons}
In this section we will review a few results on characteristic classes and instantons on a K3 manifold $M$.
These will be useful in constructing explicit examples of $N=2$ heterotic vacua.
For the most part this is standard material, with nice presentations in~\cite{Bershadsky:1996nh,Honecker:2006dt}.
First, we note that if the $\SO(d)$ structure of a manifold $X$ is reduced to $\SU(d)$, then $p_1(T_X) = 2\ch_2(T_X)$.  Thus, for $M$ we have $p_1(T_M) = -48$.\footnote{By an abuse of notation, $p_1$ will denote the differential form, the corresponding cohomology class, or its integral over the K3, as follows from the context.}

Given a vector bundle $E \to M$ with structure group a connected simple group $G_E$, we can form an associated principal $G_E$ bundle $P \to M$.  The topological classification of such bundles on compact connected four-dimensional Riemannian manifolds is discussed in the appendix of~\cite{MR658473}.  The result is that for simply connected $G_E$ the bundles $P \to M$ are classified by the first Pontryagin class.  When $G_E$ is not simply connected, one more topological invariant is needed --- a choice of a map from the classifying space to $H^2(M, \pi_1(G_E))$.  Indeed we already encountered an example of this invariant for $G_E = \SO(k)$:  the Stiefel-Whitney class $w_2(E) \in H^2(M,\Z_2)$.

The moduli space $\cM(\cA)$ of anti-self-dual (ASD) connections (when such connections exist) modulo gauge transformations has quaternionic 
dimension determined by an index computation combined with some vanishing theorems~\cite{Atiyah:1978wi}\footnote{See~\cite{Freed:1991in,Donaldson:1990bk} for an in-depth discussion of existence of ASD connections, as well as conditions when the virtual dimension computed by the index is the actual dimension of the moduli space.}
\begin{align}
\label{eq:instmod}
N_H^0 \equiv \dim_\H \cM(\cA) = -h_{\Lg} \frac{p_1(E)}{2} - \dim \Lg.
\end{align}
For ASD connections $p_1(E) < 0$;  for example, for $G = \SU(n)$ we have $p_1(E) = -2 c_2(E)$
and $N_H^0 = n c_2(E) - n^2+1$.

Unlike the more involved case of HYM connections over higher dimensional Calabi-Yau manifolds,
there is no possibility of higher obstructions:  given a smooth HYM connection over a smooth $M$,  the $N^0_H$ deformations of the connection, as well as the deformations of the Calabi-Yau metric and B-field on $M$ can all be integrated to finite deformations.

\subsection{Abelian instantons}\label{ss:abinst}
The moduli space of irreducible connections is compactified by including reducible connections.  Some of these correspond to point-like instantons and lead either to strongly coupled CFT (when the zero-size instanton is located on a smooth point in $M$)~\cite{Witten:1999fq} or important string non-perturbative effects (when the zero-size instanton is located at a singularity)~\cite{Witten:1995gx}.  The latter have been used to great effect in~\cite{Aspinwall:1997ye}.  In this work we will stick to theories where the NLSM is a good description, so we will not discuss the zero-size instantons.  However, there are plenty of reducible connections where the theory remains weakly coupled.  Perhaps the nicest example of such limiting points is provided by abelian instantons, where the structure group is reduced to $\GU(1)^m$, or equivalently, the vector bundle splits as 
$E = \oplus_a L_a$ for some
holomorphic line bundles $L_a$ on $M$.  Line bundles on $M$ are characterized by $\Pic(M) \equiv H^{(1,1)}(M,\C) \cap H^2(M,\Z)$, and for generic complex structure $\Pic(M)$ will be empty.  Let $J$ and $\Omega$ be, respectively, the K\"ahler and holomorphic (2,0) forms on $M$.  Then denoting by $\cdot$ the intersection
product $H^2(M)\times H^2(M) \to H^4(M)$ we have the familiar conditions~\cite{Aspinwall:1996mn}
\begin{align} 2! J\cdot J = \Omega \cdot \Omegab > 0, \qquad \Omega\cdot J = \Omega\cdot \Omega = 0.\end{align}
Accounting for an $\SU(2)$ rotation of $(J, \Rea \Omega, \Img \Omega)$, these specify a $58$-dimensional
family of $\SU(2)$ structures on $M$; by Yau's theorem each point in this moduli space 
determines a unique hyper-K\"ahler metric on $M$.  As is familiar, the moduli space of the corresponding (0,4)
conformal theory includes a choice of closed $B \in H^2(M,\R)$, leading to a quaternionic-K\"ahler moduli space of real dimension $80$.\footnote{This is a bit imprecise since shifting $B$ by a class in $H^2(M,2\pi\alpha' \Z)$ will leave the action invariant; in what follows we will neglect this, as well as additional discrete structure on the moduli space.  More details can be found in~\cite{Aspinwall:1996mn,MR2172498}.}

If we demand that $M$ also admits a holomorphic line bundle $L_a$ with connection $A_a$ and curvature $F_a$ then $c_1(L_a) = \ff{1}{2\pi} F_a \in \Pic(M)$; i.e. $\Omega \cdot c_1(L_a) = 0$.  If we also 
demand that the curvature $F_a$ is ASD, then $J\cdot F_a = 0$.  Thus, every linearly independent $c_1(L_a) \in \Pic(M)$ reduces the real dimension of compatible metrics by $3$.  The Green-Schwarz mechanism
leads to an additional reduction in the CFT moduli space.  This follows from~(\ref{eq:delB}) because under global gauge transformations with constant parameters $\ep_a$  the B-field shifts by $\delta B = -\ff{1}{2} \ep^a F_a$,
so that the B-field moduli, instead of residing in $H^2(M,\R)$ are actually characterized by $H^2(M,\R) / \{ \Span\{F_a\} \subset H^2(M,\R)\}.$  Thus, if $E = \oplus_{a}^k L_a$ with $k$ linearly independent classes $c_1(L_a)$, then the quaternionic dimension of the CFT moduli space is reduced by $k$.  

The left-moving current algebra is also affected by the non-trivial abelian instantons.  Very naively, one
might think that the current algebra should be the commutant of $\Lg_E \subset \so(32) ~\text{or}~ \Le_8\oplus \Le_8$; however, the gauge transforming components of the $B$-field act as St\"uckelberg fields that give masses to the $\GU(1)^k$ gauge bosons.  The spacetime interpretation of this phenomenon goes
back to~\cite{Dine:1987xk}; it has been discussed in the K3 context in, for instance,~\cite{Honecker:2006dt,Honecker:2006qz}, and more recently in the context of F-theory/heterotic compactifications on 
Calabi-Yau three-folds in~\cite{Donagi:2009ra,Anderson:2011ty}.  The worldsheet mechanism has been
recently discussed in~\cite{Melnikov:2011ez}.

\subsubsection*{Some massless spectra}
Let us describe some examples of heterotic compactifications with $8$ supercharges and
abelian instantons; in what follows we will see a very similar structure for heterotic flux vacua.  
For concreteness we work with the $\Spin(32)/\Z_2$ string.

To descirbe the line bundle $E = \oplus_{a=1}^m L_a$ in the free fermion construction we group the
$32$ fermions $\lambda^A$ into $m$ Weyl fermions $\lambda^a$ and their conjugates $\lambdab^a$
and $32-2m$ free fermions $\xi^\alpha$.  The kinetic term of the $\lambda^a$ is 
\begin{align} \frac{1}{4\pi} \lambdab^a (\pb \lambda^a + i \pb\phi^j A_{aj} \lambda^a),\end{align}
and $F_a = d A_a \in 2\pi H^2(M,\Z)$.  The anomaly cancelation conditions are then
\begin{align}
\label{eq:anomab} c_1(E) = \sum_{a} c_1(L_a) = 0\mod 2, \qquad p_1(E) = 2 \ch_2(E) = \sum_a c_1(L_a)^2 = -48.\end{align}

Assuming that we can make the corresponding NLSM weakly coupled, the naive spectrum of massless
fermions has a simple presentation~\cite{Distler:1987ee}.  This is especially true for the $\Spin(32)/\Z_2$ string since, unlike in the $\GE_8\times\GE_8$ string, all massless states arise in the (NS,R) sector.  Labeling the right-moving fermion zero modes $\psib^{\ib}$ and $\psi^i$, we take the ground state to be annihilated by
the $\psi^i$, so that the low energy states take the form
\begin{align} (\text{left-moving excitations}) \times \omega_{\ib_1\cdots \ib_k}(\phi,\phib) \psib^{\ib_1} \cdots \psib^{\ib_k} |0\ra,\end{align}
where $\omega$ belongs to an appropriate Dolbeault cohomology group.
The possible left-moving GSO-invariant left-moving excitations either involve $\lambda^A_{-1/2} \lambda^B_{-1/2}$ or $\p\phi^i$.  Ignoring the complex conjugate states to avoid double-counting, the possible states
are
\begin{align}
\begin{array}{lll}
\xi^\alpha \xi^\beta \omega|0\ra, &  \omega \in H^0(M,\cO_M), & \text{$\so(32-2m)$ gauginos} ;\\
\xi^A \lambda^a \omega|0\ra, & \omega \in H^1(M,L_a), & \text{charged hyperinos}; \\
\lambda^a\lambda^b\omega|0\ra, & \omega\in H^1(M,L_a\otimes L_b), & \text{
$\so(32-2m)$-neutral hyperinos};\\
\lambda^a\lambdab^b\omega|0\ra, ~a>b, & \omega\in H^1(M,L_a\otimes L_b^\ast), & \text{ $\so(32-2m)$-neutral hyperinos};\\
\lambda^a\lambdab^a \omega |0\ra, &\omega \in H^0(M,\cO_M) &\text{$m$ $\Lu(1)$ gauginos}; \\
\lambda^a\lambdab^b \omega |0\ra, ~a>b&\omega \in H^0(M,L_a\otimes L_b^\ast) &\text{possible additional gauginos}; \\
\p\phi^i \omega |0\ra, &\omega \in H^1(M,T^\ast), &\text{20 neutral $\KT$ hyperinos}.
\end{array}
\end{align}
As discussed above, the last two types of states mix, and only certain
linear combinations are massless.  If all $m$ classes $c_1(L_a)$ are linearly independent
in $H^2(M,\R)$, then all of the $\GU(1)^m$ gauginos are massive, and there remain $20-m$ K3 moduli.
Linear dependence will lead to enhanced gauge symmetries and additional moduli, but for simplicity 
we will stick to the case of $m$ independent classes.  In a theory with 8 supercharges we need not
worry about higher order obstructions, so that every first order deformation we find can be integrated up
to a finite deformation.  In this case we can use the index theorem to
compute the number of massless states.\footnote{As usual, we assume that we are at a generic enough
point in the moduli space such that the index theorem accurately describes the spectrum; indeed, we already assumed this in listing the relevant cohomology groups in the table.}  
The Hirzebruch-Riemann-Roch theorem for a Hermitian bundle $E$ on $M$ states that
\begin{align} \chi(E) \equiv h^0(E) -h^1(E) + h^2(E) = \int_M \ch(E) \Td(M) = 2 \rank E + \ch_2 (E),\end{align}
which for a line bundle $L$ on $M$ reduces to the familiar~\cite{Griffiths:1978pa}
\begin{align} \chi(L) = 2 + \ff{1}{2} c_1(L)^2.\end{align}
This is clearly an integer since the intersection form on $M$ is even.  Applying this to the states
above, we find that the massless spectrum consists of the $\so(32-2m)$ vector multiplets, 
$N^+_H$ hypers in the fundamental representation of $\so(32-2m)$ and $N_H^0$ neutral hypers with
\begin{align}
N^0_H & = 20-m -\sum_{a>b} \left[\chi(L_a \otimes L_b) +\chi(L_a\otimes L^\ast_b) \right]  = 20-m + (48-2m)(m-1), \nonumber\\
N^+_H & = [-\sum_{a} \chi(L_a)] \times(\rep{32-2m}) = (24-2m)\times(\rep{32-2m}),
\end{align}
where we used $\sum_a c_1(L_a)^2  = -48$.  We can see that $N_V - N_H = 244$ as 
is appropriate for a perturbative heterotic spectrum in $6$ dimensions.

This six-dimensional theory can be compactified further
on $T^2$; by turning on Wilson lines for the gauge fields along the torus, i.e. the vector 
multiplet moduli in the four-dimensional
theory, we can break $\so(32-2m) \to \Lu(1)^{\oplus (16-m)}$; at a sufficiently generic point
this also lifts all of the charged matter hypers. Combining the resulting massless states with the 
three $\GU(1)$ vector multiplets due to $T^2$, we find a four-dimensional theory with
\begin{align} N_V = 19-m, \qquad N_H = 20-m +2 (24-m)(m-1). \end{align}
The same progression of $N_V(m)$ and $N_H(m)$ can be obtained by a slight
variation of the four-dimensional construction.  At a fixed $m$ we can go to the origin of the
Coulomb branch, recovering $\so(32-2m)$ gauge group and corresponding charged hypers;
we can then partially Higgs the theory from $\so(32-2m) \to \so(30-2m)$ and go on the
Coulomb branch of $\so(30-2m)$.  The resulting change of spectrum is exactly the same
as that obtained by changing $m \to m+1$.

Interpreting these spectra in terms of potential IIA duals leads to a set of 
Calabi-Yau manifolds $Y_m$ , $m=1,\ldots, 12$, with Hodge numbers
\begin{align}
\label{eq:abinst1}
(h^{1,1},h^{1,2}) &\in \{(18,18),~(17,61),~(16,100),~(15,135),~(14,166),~(13,193),~(12,216) ,\nonumber\\
&\qquad (11,235),~(10,250),~(9,261),~(8,268),~(7,271)\}.
\end{align}
All of these are realized by known constructions.\footnote{Our Calabi-Yau data mining was greatly expedited by the database of known Calabi-Yau constructions maintained by B.~Jurke at~\url{http://cyexplorer.benjaminjurke.net}.}

\subsection{Criteria for smooth $M$} \label{ss:smooth}
The list of models above terminates at $m=12$.  A reason to distrust the results for $m>12$
is that $N^+_H$ becomes negative; however, in our geometric description there is a more direct
way of identifying a problem.  Recall that a K3 $M$ is singular if and only if $\Pic(M)$ contains
a $-2$ curve of zero size~\cite{Aspinwall:1996mn,Barth:2004ne}.  Equivalently, $M$ is singular
if and only if it admits an abelian instanton with $c_1(L)^2 = -2$. Since the K3 intersection
lattice is even, an abelian instanton, since it is anti-self-dual, satisfies $c_1(L)^2 \le -2$; therefore for $m >12$ 
$M$ is necessarily singular, with a point-like instanton supported at the singularity.  This sort of
singularity in the CFT is outside of the domain of string perturbation theory, and its resolution
is often accompanied by enhanced gauge symmetries and extra matter states.

For $m\le 12$ it is possible to realize the instanton configuration on a smooth $M$.  Consider
$M$ to be the Kummer surface, i.e. $T^4/\Z_2$ blown up at the $16$ singular points, with exceptional
divisors $E_i$, $i=0,\ldots,15$.  These have self intersection $E_i \cdot E_j = -2\delta_{ij}$. 
Consider the $12$ linearly independent divisors
\begin{align}
\label{eq:Divs}
D_1 & = E_0 - E_5, & D_5 & = E_8 - E_0, &  D_9 & = E_7 - E_6,\nonumber\\
D_2 & = E_1 - E_5, & D_6 & = E_9 - E_2, &  D_{10} & = E_7-E_{10},\nonumber\\
D_3 & = E_2 - E_{15}, & D_7 & = E_{10} - E_{4}, &D_{11} & = E_{13} - E_{12},\nonumber\\
D_4 & = E_3 - E_{15}, &D_8 & = E_{11} - E_8, &D_{12} & = E_{13}- E_{14}.
\end{align}
Evidently the corresponding line bundles $L_a = [D_a]$ have $c_1(L_a)^2 = -4$.  In addition, $\sum_a c_1(L_a) = 0 \mod 2$, so that~(\ref{eq:anomab}) is satisfied.  The last point uses
the fact (see appendix B of~\cite{Kumar:2009zc} for a clear presentation of a nice basis of $H^2(M,\Z)$) that
the classes
\begin{align}
I_1  & = \ff{1}{2} (E_0 + E_1 + E_2 + E_3 + E_8 + E_9 + E_{10} + E_{11}), \nonumber\\
I_2  & = \ff{1}{2} (E_0 + E_2 + E_4 + E_6 + E_8 + E_{10}+E_{12}+E_{14})
\end{align}
are in $H^2(M,\Z)$.  Finally, each $L_a$ will admit an ASD connection if we take
all of the exceptional divisors to have a common size $J(E_a) = j$.  By taking linear combinations
of the $D_a$ we can produce all of the $1\le m \le 12$ examples.  For instance, we obtain the $m=1$ example by taking $D = \sum_{a=1}^{12} D_a$. 

\section{Some potential IIA duals of heterotic flux vacua} \label{s:duals}
Having discussed the heterotic worldsheet theory at some length, we now turn to their potential type II
duals.  A generic heterotic vacuum will not have a weakly-coupled type II dual, and to describe its
non-perturbative features requires some more general formalism in the spirit of F-theory.  However,
there is a non-trivial class of string vacua that include weakly coupled type II and heterotic limiting points
in the moduli space.  Identifying these tractable dual pairs is important since such vacua offer a nice
laboratory for studying string non-perturbative effects.  How do heterotic flux vacua fit into this class of
theories?  

To frame the discussion let us first recall some powerful constraints on possible type II duals
of perturbative heterotic vacua with eight supercharges. Quite early on it
was appreciated that K3-fibered and elliptically-fibered Calabi-Yau three folds should play
a special role in the
duality~\cite{Kachru:1995wm,Ferrara:1995yx,Klemm:1995tj}.  The K3-fibration structure has
a particularly elegant explanation from the perspective of the heterotic
conformal field theory~\cite{Aspinwall:1995vk}.
  In the weak coupling
limit, the special K\"ahler geometry of the vector moduli space has a universal form determined
by a cubic prepotential~\cite{deWit:1995zg}:
\begin{align} F_0 = - \gamma_{ij} T^i T^j S +F_0^1(T)+\ldots, \qquad \gamma = \diag (+,-,\ldots,-),\end{align}
where $S$ is the axio-dilaton modulus, the $T^i$ denote the remaining vector moduli, the
$F_0^1$ is the one-loop correction, and the $\ldots$ signify string non-perturbative corrections.
In the same notation, the prepotential $F_1$  --- the coefficient of the $R^2$ coupling in the effective
four-dimensional theory --- has a universal form
\begin{align} F_1 = 24 S + F^1_1(T)+ \ldots.\end{align}

If we suppose that the type IIA dual of this weakly coupled limit corresponds to a large radius phase
of a  compactification on a smooth Calabi-Yau 3-fold $Y$, then we can compare the above structure to the type II 
results.  In this case, the structure of the vector moduli space is completely determined by the A-model topological string associated to $Y$, and neglecting worldsheet and perturbative corrections, the prepotentials are given by
\begin{align} F_0 = - \frac{i}{6} D_A \cdot D_B \cdot D_C T^A T^B T^C + \ldots, \qquad
    F_1 = -\frac{4\pi i}{12} D_A \cdot c_2(Y)  T^A + \ldots.\end{align}
Here  $A=0,\ldots, h^{1,1}(Y)-1$, $\{D_A\}$ is a basis for the divisor classes on $Y$, and $\cdot$ denotes
divisor intersection.
Comparing this structure to the heterotic result leads to constraints on the geometry of $Y$:  there 
exists a distinguished divisor $D_0$ such that $D_0^2\cdot D_A = 0$ for all $A$ and
$D_0\cdot c_2(Y)  = 24$.  In addition, it is argued in~\cite{Aspinwall:1995vk,Aspinwall:1996mn} that
convergence of worldsheet instanton sums requires $D_0$ to be a numerically
effective (NEF) divisor, i.e. for any algebraic curve $C$ in $Y$, $ D_0 \cdot C \ge 0$.  These conditions 
are sufficient to show that $Y$ is a K3 fibration, with $D_0$ being the class of the generic 
fiber~\cite{MR1228584}.

The F-theory perspective identifies another important fibration structure in type II Calabi-Yau compactifications:  $Y$ can be elliptically fibered with section.  The conditions on divisors for the existence of such a fibration were studied in~\cite{MR1228584} and reviewed in~\cite{Morrison:1996na}:  there exists a NEF divisor $D_1$ (the class of the section) with $D_1^3 = 0$ and $D_1^2 \cdot D_2 =1$ for some other divisor $D_2$.  The K3 and elliptic fibrations are compatible if $D_0 \cdot D_1^2  = 0$.  Since $Y$ is K\"ahler, and the K\"ahler class is positive, a  manifold with such a structure necessarily has $h^{1,1}(Y) \ge 3$.

The relevance of this compatible elliptic fibration for heterotic/type II duality is a consequence of fiberwise application of the duality between F-theory on an elliptically fibered Calabi-Yau three-fold and heterotic compactification on a K3~\cite{Morrison:1996na}:  if the
heterotic description has a limit where the $T^2$ can be taken to be arbitrarily large, then $Y$ admits a compatible elliptic fibration with at least one section~(see~\cite{Aspinwall:2000fd}, in particular proposition~10).\footnote{In N=2 Calabi-Yau compactifications of type II theories the elliptic fibration ensures that the theory can be lifted to a supersymmetric theory in six dimensions~\cite{Ferrara:1996wv}.}  In table~\ref{table:fibs}
we provide some examples of three-folds, listing their Hodge numbers and note the existence of a K3 fibration ($\Pi_{\KT}$), elliptic fibration with section ($\Pi_{E}$) and their compatibility (C); many additional examples can
be found in~\cite{Klemm:1995tj,Avram:1996pj,Klemm:2004km}. 
\begin{table}[t]
\begin{center}
\begin{tabular}{cccccccc}
hypersurface			&$h^{11}$		&$h^{12}$		&$\Pi_{\KT}$	&$\Pi_{E}$	&C	 \\ 
$Y_{18} \subset \P^4_{11169}$ &$2$		&$272$		&\cross		&\tick		&-- 	 \\
$Y_{24} \subset \P^4_{1128,12}$ &$3$	&$243$		&\tick		&\tick		&\tick \\
$Y_{12} \subset \P^4_{11226}$ &$2$	&$128$		&\tick		&\cross		&-- 	 \\
$Y_8\subset \P^4_{11222}$	&$2$		&$86$		&\tick		&\cross		&--	 
\end{tabular}
\caption{Examples of fibration structures in three-folds.}
\label{table:fibs}
\end{center}
\end{table}

With these facts in hand, we now see that there is a natural guess for weakly coupled duals to heterotic flux vacua.
Since the K3-fibration structure follows from properties of the heterotic conformal field theory, we still expect the dual geometry $Y$ to be K3-fibered; however, we have also seen that in a typical heterotic flux vacuum the torus geometry is fixed, and there is no six-dimensional decompactification limit.  Thus, we can expect $Y$ to lack a compatible elliptic fibration with section.  Conversely, given a type II vacuum based on a K3-fibered $Y$ without an elliptic fibration a perturbative heterotic dual, if it exists, must necessarily be a heterotic flux vacuum.\footnote{This issue is a little bit clouded by T-dual descriptions of principal torus bundle target spaces; we will discuss this in more detail below.}

For instance, from our discussion it is clear that the large radius limit of the $Y_{18}$ hypersurface cannot be dual
to a weakly coupled heterotic string, while the remaining examples can have weakly coupled duals.
Indeed, the duals of $Y_{24}$ and $Y_{12}$ were proposed in~\cite{Kachru:1995wm} and subjected to
further tests in~\cite{Kaplunovsky:1995tm}.  The last example is familiar in the context 
of mirror symmetry~\cite{Candelas:1993dm,Hosono:1993qy}; it and $Y_{12}$ are the only known examples of a 
two-parameter K3-fibered Calabi-Yau three-fold hypersurface in a toric variety.  Since neither example has
an elliptic fibration, we do not expect the CFT of the heterotic dual to consist of decoupled
$T^2$ and $\KT$ components.  We will construct some new potential heterotic duals for interesting K3-fibered 
$Y$ with with low Hodge numbers below.  First, however, we will examine some abelian instanton examples
that are closely related to those in section~\ref{ss:abinst}.

\subsection{Abelian instanton examples}
Let $n \in \{1,2\}$ label the number of non-trivially fibered cycles of $T^2$, with corresponding line bundles $\Lt_{I}$.\footnote{The $n=0$ examples, where the torus is trivially fibered, were discussed in the previous section.}  We consider again the $\Spin(32)/\Z_2$ string with bundle $E = \oplus_{a=1}^m L_a$.  For simplicity we will take $\cG^\ast_{IJ} = \ba \delta_{IJ}$, $b = 0$,  $\bca_I \cFh = 0$. Note that this choice of $\cG^\ast$ with zero Wilson lines does not lead to any enhanced gauge symmetry unless $\ba = \alpha'$: the $T^2$ is a square torus with equal radii $\sqrt{\alpha'/2}$, and in our conventions the self-dual radius is $\sqrt{\alpha'}$.  As a final simplification, we will also assume that the line bundles $\Lt_{I}$ and $L_a$ correspond to $n+m$ linearly independent classes in $H^2(M,\Z)$.

The flux quantization conditions of section~\ref{ss:fluxquantization}, and in particular~(\ref{eq:finalfluxquantization}), will be satisfied if and only if
\begin{align}
\label{eq:fluxquantization}
\frac{\ba}{\alpha'} c_1(\Lt_I) \in H^2(M,\Z)~.
\end{align}
We need the line bundles $L_a$ to satisfy the global anomaly constraint
\begin{align}
\label{eq:globalanom}
\sum_{a=1}^m c_1(L_a) \in H^2(M,2\Z)~.
\end{align}
\noindent\textit{These conditions were not imposed correctly in a previous version of the paper.  The corrections made below lead to a modification of the proposed Hodge numbers for the dual type II geometries.  In particular, now every proposed Hodge number pair is realized by a known Calabi-Yau geometry.  Furthermore, a number of vacua with $N_V = 2$ constructed in section 4.3 in the earlier version of the paper turn out to be spurious.}\\[2mm]

We will now explore the possible values of torus area $\ba$ as well as choices of ASD line bundles $\Lt_I$ and $L_a$ that are consistent with the Bianchi identity,~(\ref{eq:fluxquantization}), and~(\ref{eq:globalanom}), and a smooth K3 geometry.   First, we see that  we need
\begin{align}
\ba = \frac{p\alpha'}{q}~,
\end{align}
where $p,q$ are relatively prime positive integers.  When this holds~(\ref{eq:fluxquantization}) is satisfied if and only if there exist line bundles $\cL_{I}$ such that
\begin{align}
\Lt_I = \cL_{I}^{\otimes q}~.
\end{align}
In this case the Bianchi identity is recast as
\begin{align}
\label{eq:Bianchipq}
2pq \sum_{I=1}^n c_1(\cL_I)^2+ \sum_{a=1}^m c_1(L_a)^2& = -48~.
\end{align}
Provided we can satisfy the conditions of quantization and supersymmetry on a smooth geometry, then following the discussion in the previous section, we obtain a theory with an unbroken gauge algebra $\Lu(1)^{\oplus (n-3)}\oplus\so(32-2m)$, and the matter spectrum consists of 
\begin{align} 
N'^0_H &= 20-m-n + (48+k-2m)(m-1), \nonumber\\
N'^+_H &= (24-2m+k/2) \times \rep{(32-2m)}.
\end{align}
This uses the index theorem, the results on flux quantization, and $k$ is the contribution to~(\ref{eq:Bianchipq}) from the torus fibration:
\begin{align}
k = 2pq \sum_{I=1}^n c_1(\cL_I)^2~.
\end{align}
The hypermultiplets are neutral under $\Lu(1)^{\oplus(3-n)}$.  At a generic point on the
Coulomb branch the gauge group is broken to a Cartan subgroup $\GU(1)^{19-n-m}$, and all charged hypermultiplets become 
massive; a type IIA interpretation therefore corresponds to a Calabi-Yau manifold with Hodge numbers
\begin{align}
\label{eq:possiblehodge}
h^{1,1} & = 19-n-m~,&
h^{1,2} & = 19-m-n + (48+k-2m)(m-1)~.
\end{align}

We will now characterize the $k$ and $m$ that are realized by supersymmetric configurations on smooth $M$ and with properly quantized flux.

Let $\cD_a$ label the divisor associated to the gauge line bundle $L_a$, i.e. $L_a = [\cD_a]$.  If $m\neq 0$ we require these to satisfy
\begin{align}
\label{eq:gaugebundleconstraints}
J\cdot \cD_a & = 0~,&
\cD_a\cdot \cD_a &\le -4~,&
\sum_{a=1}^m \cD_a \cdot \cD_a &= -(48+k)~, &
\sum_{a=1}^m \cD_a & = 2 D~,
\end{align}
where $D \in H^2(M,\Z)$.
The first two conditions are required for $L_a$ to be an abelian instanton bundle on a smooth K3; the second one is the Bianchi identity, and the last one is the global anomaly constraint.  Clearly we must have
\begin{align}
\label{eq:mequation}
m \le \frac{48+k}{4}~.
\end{align}
We will now argue that $k\in\{-16,-24,-32,-48\}$.

Since the $\cD_a$ are orthogonal to $J$ and $\Omega$, the intersection pairing on $H^2(M,\R)$ induces a negative-definite pairing on the $m$-dimensional subspace of $H^2(M,\R)$ generated by the $\cD_a$, and therefore we have by the triangle inequality
\begin{align}
-(48+k) = \sum_{a=1}^m \cD_a \cdot \cD_a \le 4 D\cdot D.
\end{align}
But $D$ is an integral anti-self-dual divisor, and on a smooth K3 we must have $D\cdot D \le -4$.  Thus it is not possible to realize $k = -40$ or $k=-36$.

On the other hand, constraints from the fibration lead to an upper bound on $k$.  To avoid the self-dual radius we need to have $pq >1$, and for a smooth K3 we need $c_1(\cL_I)^2 \le -4$.
Therefore, if $n=1$ (i.e. one fibered circle), then $k\le -16$, while for $n=2$ we need $k \le -32$.  Since $c_1(\cL_I)^2$ is also even, the possible $k$ values are very limited.  For $n=1$ we obtain
\begin{align}
k &= -16 \iff (pq,c_1(\cL_1)^2) = (2,-4)~, \nonumber\\
k &=-24 \iff (pq,c_1(\cL_1)^2) \in \{ (2,-6)~,(3,-4)\}~,\nonumber\\
k &=-32 \iff (pq,c_1(\cL_1)^2) \in \{ (2,-8)~,(4,-4)\}~,\nonumber\\
k &=-48 \iff (pq,c_1(\cL_1)^2) \in \{ (2,-12)~,(3,-8)~,(4,-6)~,(6,-4)\}~.\nonumber\\
\end{align}
For $n = 2$ the possibilities are even more meager:
\begin{align}
k &=-32 \iff (pq,c_1(\cL_1)^2,c_1(\cL_2)^2) = (2,-4,-4)~,\nonumber\\
k &=-48 \iff (pq,c_1(\cL_1)^2,c_1(\cL_2)^2) \in \{ (2,-4,-8)~,(2,-6,-6)~,(3,-4,-4)\}~.\nonumber\\
\end{align}
Each $k$ value has a realization with $n$ $-4$--classes and some $pq$.  

Using the example of the Kummer surface discussed above we found explicit realizations with $n+m$ linearly independent ASD line bundles for every $k \in \{-16,-24,-32,-48\}$ and $0\le m\le (48+k)/4$.  Each of these leads to a heterotic flux vacuum and therefore conjecture IIA dual with Hodge numbers in~(\ref{eq:possiblehodge}).  Scanning through the integers, we find that for $n=1$ the $(h^{1,1},h^{1,2})$ are in the following list:
\begin{align}
\{(10, 122)~,&& (11, 119)~,&& (12, 72)~,&& (12, 112)~,&& (13, 69)~,&& (13, 101)~,&& \underline{(14, 38)}~, \nonumber\\
   (14, 62)~,&& (14, 86)~,&& \underline{(15, 35)}~,&& (15, 51)~,&& (15, 67)~, &&\underline{(16, 28)}~, &&(16, 36)~,  \nonumber\\
   (16, 44)~,&& \underline{(17, 17)}~,&& (18, 18)\}~.
\end{align}
For $n=2$ we obtain a shorter list, which only consists of the underlined pairs.  All $17$ pairs of Hodge numbers appear in the Kreuzer--Skarke database~of Calabi-Yau three-folds~\cite{Kreuzer:2002uu}, and all but two ( the pairs $(10,122)$ and $(12,112)$) are realized by K3-fibered three-folds~\cite{Avram:1996pj}.

It would be interesting to determine which of the matched Hodge pairs have known realizations that admit K3 fibrations and do not admit elliptic fibrations.  We leave this for future investigation and instead turn to some examples with $h^{1,1} = 2$.

\subsection{IIA/heterotic dual pairs with two vector multiplets} \label{ss:twovectors}
One of the earliest examples of IIA/heterotic duality was obtained as follows~\cite{Kachru:1995wm}.  The $\GE_8\times\GE_8$ heterotic string was compactified to $d=8$ on a $T^2$ with $\tau = \rho $, leading to an enhanced gauge symmetry $\GU(1)^2\times \SU(2)\times\GE_8\times\GE_8$.
This was then compactified further on a $\KT$ manifold $M$ with instantons
\begin{align} \SU(2)_{c_2 = 4} \times \SU(2)_{c_2 = 10} \times \SU(2)_{c_2 = 10} \subset \SU(2)\times\GE_8\times\GE_8,\end{align}
leaving a four-dimensional theory with gauge group $\GU(1)^2 \times \GE_7\times \GE_7$ with $3$ $\rep{56}$s for each $\GE_7$.  Higgsing the $\GE_7\times \GE_7$ leads to $N_V = 2$ and $N_H = 129$, suggesting a dual Calabi-Yau geometry with $h^{1,1} = 2$ and $h^{1,2} = 128$.  A comparison of the vector moduli space geometry in the two descriptions~\cite{Kaplunovsky:1995tm} offered a compelling test of the duality.

It is instructive to carry out the same construction with more general values of instanton numbers
\begin{align} \SU(2)_{c_2 = k_0} \times \SU(2)_{c_2 = k_1} \times\SU(2)_{c_2 = k_2} \subset \SU(2)\times\GE_8\times\GE_8, \qquad k_0+k_1+k_2 = 24.\end{align}
In order to have irreducible $\SU(2)$ connections we require $k_{0,1,2} \ge 2$, in which case the dimension of the moduli space is given by~(\ref{eq:instmod}).  Using the decomposition $\GE_8 \to \SU(2)\times\GE_7$, under which
\begin{align} \rep{248} = (\rep{3},\rep{1}) + (\rep{2},\rep{56}) + (\rep{1},\rep{133}),\end{align}
and the index theorem, we see that the $\GE_7\times\GE_7$-charged matter spectrum consists of\footnote{Since $\rep{56}$ is pseudo-real, it is possible to have half-hypermultiplets;  since $\pi_4(\GE_7) = 0$ an odd number of half-hypermultiplets does not lead to a global anomaly.} 
\begin{align} (\ff{1}{2} k_1 -2) \times (\rep{56},\rep{1}) + (\ff{1}{2} k_2 -2) \times (\rep{1},\rep{56}).\end{align}
A necessary requirement to completely Higgs $\GE_7\times\GE_7$ is $k_{1,2} \ge 9$.  If we assume that complete Higgsing is possible for $k_{1,2} \ge 9$, then on that Higgs branch we obtain
a theory with $G= \GU(1)^2$ and a number of possibilities for the number of  $G$-neutral hypermultiplets $N_H^0$:
\begin{align}
\begin{array}{lcccc}
(k_1,k_2)					&~	& N_H^0  	&~	& \text{list?} \\[0.1cm]
(9,9)						&~	&73		&~	& \ttick \\
(9,10)					&~	&101		&~	& \ttick \\
(9,11);(10,10)				&~	&129		&~	& \ttick \\
(9,12);(10,11)				&~	&156		&~	& \tcross \\
(9,13);(10,12);(11,11)		&~	&184		&~	& \tcross
\end{array}
\end{align}
The middle row with $k_1=k_2=10$ is the example discussed above.  
What of the first two rows?  The corresponding Calabi-Yau three-folds exist, and they are indeed
K3-fibered.  They were constructed as co-dimension $2$ complete intersections in toric
varieties~\cite{Klemm:2004km}.  There are no known examples of Calabi-Yau three-folds that
could realize the spectra of the last two rows.

Our assumption about complete Higgsing may be too naive in the $k=9$ case.  The trouble is that sequential Higgsing $G\to G_1 \to G_2 \to \cdots \to 1$,
where at each step a vacuum expectation value is assigned to a single irreducible representation, does not lead to complete Higgsing.\footnote{Sequential chains have been extensively studied in the context of type II/heterotic duality, with successive gaugings often finding a combinatorial interpretation in a ``chain'' of reflexive polytopes, e.g.~\cite{Aldazabal:1995yw,Candelas:1996su}.}   As we discuss in appendix~\ref{app:higgs}, there is no trouble in choosing expectation values of the hypermultiplets so that the stabilizer subgroup is trivial; however, showing that such a configuration is indeed a supersymmetric vacuum is fairly involved.  We have not been able to find a solution, and furthermore, there is a simple argument that such configurations cannot be obtained at the level of supergravity.\footnote{The unbroken gauge group arises as the commutant of the gauge connection valued in $H \subset \GE_8$.  $\GE_6$ is the largest subgroup of $\GE_8$ that admits irreducible connections with $k=9$, so $H$ is at most $\GE_6$, leading to a spacetime gauge group of at least $\SU(3)$.}  However, as suggested in~\cite{Duff:1996rs}, full Higgsing may nevertheless be possible at some special locus in the moduli space.  We find it encouraging that there exist Calabi-Yau manifolds as potential duals for $k=9$ theories with full Higgsing.   It would be interesting to explore this in more detail and determine whether the ``matching'' Calabi-Yau manifolds are just a fluke, or whether complete non-sequential Higgsing is possible for $k=9$ at least at some appropriate locus in the moduli space.

\subsection{T-duality orbits}
We end our discussion of flux vacua and their duals with a comment on T-duality.  Heterotic compactifications on principal $T^n$ bundles admit a rich structure of T-dual orbits, which include physically equivalent vacua with topologically different backgrounds.  For instance, it is possible to ``trade'' a fibered torus direction for an abelian instanton embedded in the gauge group~\cite{Evslin:2008zm}.

Despite this large equivalence, it is important to keep in mind that there are non-trivial restrictions on possible T-dual pairs.  For instance, consider the T-duality orbit of a $T^2\times\KT$ compactification.  The perturbative gauge symmetry of the resulting four-dimensional vacuum necessarily has rank $r\ge 3$.  Since T-duality is a symmetry of the conformal field theory, every compactification on the T-duality orbit will have the same gauge group.  So, a heterotic flux vacuum with $r < 3$ cannot be on a T-duality orbit of a theory with a trivial fibration.  Of course theories can still be related by motion in the moduli space; however, that goes beyond considerations of T-duality orbits.
%
%

\section{Fibered WZW models with (0,2)+(0,4) supersymmetry} 
\label{s:WZW}
In this section we return to consider the $N_V = 2$, $N_H = 129$ example of~\cite{Kachru:1995wm}.  Our goal is
to demonstrate that the heterotic description can be thought of as a flux vacuum, where the toroidal degrees of freedom are fibered over a K3 base $M$.  The idea is simple:  we present the torus with $\tau=\rho$ as a WZW model and then construct the (0,2)+(0,4) fibration over a K3 $M$ by gauging the left-moving $\SU(2)$ symmetry of the WZW theory.   

This is of course not a new idea.  Gauged WZW models~\cite{Gawedzki:1988hq,Witten:1991mm} have been used to construct examples of (0,2)-preserving vacua~\cite{Berglund:1995dv}.
The construction of heterotic flux vacua in this fashion was exploited in~\cite{Adams:2009av}, where a gauged WZW model was coupled to a gauged linear sigma model description of the base.  The novelty of our presentation of the fibration over the NLSM is the manifest (0,2)+(0,4) worldsheet supersymmetry.  

\subsection{WZW models with (0,1) supersymmetry}
To construct the gauge-invariant action
the worldsheet $\Sigma$ is presented as a boundary
of a three-manifold $N$:  $\p N = \Sigma$;  we fix a Lie group $G$ with Lie algebra $\Lg$, a representation $\rho:  G \to \GL(V_\rho)$, denote maps $\Sigma \to \rho(G)$ by $\sg$ and their
extensions to $N$ by $\sgt$; the associated Maurer-Cartan form pulled back to $\Sigma$ ($N$)
is denoted $\bomega$ ($\bomegat$); $\bomega = \sg^{-1} d\sg$.   Finally, we introduce a set of worldsheet fermions $\chi \in \rho(\Lg) \otimes \Kb_\Sigma^{1/2}$.

The level $k\in \Z_{\ge 0}$ (0,1) supersymmetric WZW action is~\cite{Rohm:1984ix,Witten:1991mk}
\begin{align}
S_G = \frac{k}{4\pi} \int_\Sigma d^2 z \left[ \tr_\rho \{\p \sg^{-1} \pb\sg\} - \tr_\rho\{\chi \p \chi\} \right]
- \frac{i k}{12 \pi} \int_N \tr_\rho\{\bomegat^3\}.
\end{align}
The representation $\rho$ is the smallest representation for which $e^{-S_G}$ is independent of the choice of $N$ for any integer $k$.  For instance, for $G=\SU(n)$ $\rho$ is the fundamental representation.  In what follows we will drop the representation label $\rho$.

Under variations $\delta \sg$ and $\delta\chi$, the change in the action is
\begin{align}
\label{eq:WZWvar}
\delta S_G = \frac{k}{2\pi} \int  d^2z \left[ \tr\{\sg^{-1}\delta\sg \p\bomega_{\zb}\} - \tr\{\delta\chi \p \chi\}\right].
\end{align}
Defining 
\begin{align}\omegab \equiv \bomega_{\zb} = \sg^{-1} \pb \sg, \qquad
\omega = \p\sg\sg^{-1} = \sg \bomega_z \sg^{-1},\end{align}
and using the identity $\pb \omega = \sg \p\omegab \sg^{-1}$, we
find the equations of motion
\begin{align}
\p \omegab  = 0, \qquad \pb \omega = 0,  \qquad \p\chi = 0.
\end{align}
The action $S_G$ is invariant under the (0,1) supersymmetry
\begin{align}
i\bQ_1 \cdot \sg =  \sg \chi, \qquad i\bQ_1 \cdot \chi = -(\omegab + \chi\chi).
\end{align}
We wish to couple this theory to the base NLSM for a K3 $M$ with action\footnote{In this section latin indices are the coordinate indices on $M$.  We will ignore the left-moving fermions as they play no essential role in the fibration.}
\begin{align}
S_\tbase &= \frac{1}{2\pi\alpha'} \int d^2z \left[ (g_{\mu\nu} +B_{\mu\nu}) \p \phi^\mu \pb\phi^\nu + 
g_{\mu\nu} \psi^\mu\p\psi^\nu + \p \phi^\lambda \psi^\mu\psi^\nu (\Gamma_{\mu\lambda\nu}-\ff{1}{2} dB_{\mu\lambda\nu}) \right].
\end{align}
This is invariant under $i\bQ_1 \cdot \phi^\mu =  \psi^\mu$ and $i\bQ_1 \cdot \psi^\mu = -\pb \phi^\mu$.

\subsection{The fibration}
The currents $\omegab$ and $\omega$ correspond to the chiral symmetries $\delta \sg = U(z) \sg + \sg V(\zb)$, where $U, V \in \Lg$.
The fibration is achieved by demanding that the total action is invariant under $\delta \sg = U \sg$, where $U$ is the pull-back to the worldsheet of a map $M \to \Lg$.  This requires the introduction of a $\Lg$-valued gauge field $A$ with $\delta_U A = -d U -\CO{A}{U}$.  In what follows, we will use a short-hand to denote various pull-backs of $A$:
\begin{align} A_{z} \equiv A_\mu \p\phi^\mu, \qquad
A_{\zb} \equiv A_\mu \pb\phi^\mu, \qquad
A_{\psi} \equiv A_\mu \psi^\mu.\end{align}

\subsubsection*{Gauge invariance of the bosonic theory}
The first step in constructing a gauge-invariant theory is to introduce
the minimal coupling $\omega A_{\zb}$ to cancel 
\begin{align}\delta_U S_G = \frac{k}{2\pi} \int d^2 z \tr\{U \pb \omega\}.\end{align}
The resulting action is still not gauge-invariant, but there is a unique coupling quadratic in $A$ such that $\delta_U (S_G+S_A)$ takes
a canonical form~\cite{Witten:1991mk}.  Namely, we take
\begin{align}S_A^\tbos &=- \frac{k}{4\pi}\int d^2z~ \tr\{ A_z A_{\zb} + 2 \omega A_{\zb}\}, \qquad \text{so that} \\
\delta_U S_A^\tbos & = - \delta_U S_G^\tbos + \frac{k}{4\pi} \int d^2 z \tr\{U dA_{\mu\nu} \} \p\phi^\mu\pb\phi^\nu.
\end{align}
The last term can be canceled by a transformation of the $B$-field:
\begin{align}
\delta_U B = -\frac{\alpha' k}{2} \tr\{U dA \},
\end{align}
leading to a gauge-invariant three-form 
\begin{align}
\cH \equiv dB - \frac{\alpha' k}{2} \CS_3(A).
\end{align}
Note that here the shift $dB \to \cH$ arises at the level of the classical action.  Including the one-loop contributions we described above will shift $\cH$ by $\CS_3(\cS^+)$ and $\CS_3(\cA)$, but we will concentrate on the classical terms due to gauging the WZW symmetry.

\subsubsection*{A supersymmetric fibration}
It is possible to extend the construction to maintain (0,1) supersymmetry.  It turns out that supersymmetry requires us to postulate gauge transformations of the $\chi$:  $\delta_U \chi =  \sg^{-1} dU_{\mu} \sg \psi^\mu$, and the action takes a simple form when written in terms of the gauge-invariant fermions $\cX \equiv \chi +\sg^{-1} A_\psi \sg$.\footnote{These might with good reason remind the reader of the gauge-invariant $\Psi^I = \psi^I + A^I_i\psi^i$ we met in the torus fibration.}  The supersymmetry transformations, when written in terms of $\cX$ are a bit more complicated:
\begin{align}
i\bQ_1 \cdot \sg = \sg \cX  - A_\psi\sg, \qquad
i\bQ_1 \cdot \cX = -(\omegab+\cX\cX + \sg^{-1} (A_{\zb} - \ff{1}{2} F_{\mu\nu} \psi^\mu\psi^\nu) \sg),
\end{align}
where $F = dA + A^2$.  The full supersymmetric fibered action is
then a sum of three terms:
\begin{align}
S_G & =
\frac{k}{4\pi} \int_\Sigma d^2 z \left[ \tr \{\p \sg^{-1} \pb\sg\} - \tr\{\cX \p \cX\} \right]
- \frac{i k}{12 \pi} \int_N \tr\{\bomegat^3\}, \nonumber\\
S_\tbase & = 
\frac{1}{2\pi\alpha'} \int d^2z  \left[ (g_{\mu\nu} +B_{\mu\nu}) \p \phi^\mu \pb\phi^\nu + 
g_{\mu\nu} \psi^\mu\p\psi^\nu + \p \phi^\lambda \psi^\mu\psi^\nu (\Gamma_{\mu\lambda\nu}-\ff{1}{2} \cH_{\mu\lambda\nu}) \right], \nonumber\\
S_A & =- \frac{k}{4\pi}\int d^2z \tr\{ A_z A_{\zb} + 2 \omega (A_{\zb} -\ff{1}{2} F_{\mu\nu} \psi^\mu\psi^\nu) - A_z F_{\mu\nu} \psi^\mu\psi^\nu\}.
\end{align}
All the fermionic terms are explicitly gauge-invariant except for the term proportional to 
$\tr\{(\omega + A_z) F_{\mu\nu}\}$; it is not hard to show that it too is gauge-invariant.

\subsubsection*{Projection of the right-moving fermions}
The degrees of freedom of the fibered WZW theory are not quite appropriate for our heterotic considerations:  there are too many right-moving fermions $\cX$.  The left and right central charges of the (0,1) WZW theory are
\begin{align} c = \frac{k \dim \Lg}{k+ h_{\Lg}}, \qquad \cb = c + \frac{\dim \Lg}{2}.\end{align}
For our application we need a level $1$ $\Lg = \su(2)\oplus\Lu(1)$ current algebra with $(c,\cb) = (2,3)$.  To obtain the correct theory the
$\cX$ should be valued in the Cartan subalgebra $\LLh\subset \Lg$.

To carry out this reduction of degrees of freedom in a supersymmetric fashion, we pick a projector $\Pi_{\LLh} : \Lg \to \LLh$ satisfying
\begin{align} \tr\{x \Pi_{\LLh} (y)\} = \tr\{y \Pi_{\LLh}(x)\} \qquad \text{for all}\qquad
x,y \in \Lg.\end{align}
By construction $\Pi_{\LLh}(\cX) = \cX$, and we form a modified  supercharge $\bQ_1^{\text{new}} = \bQ_1^{\text{old}}$ on $\phi,\psi$ and $\sg$, while
\begin{align}
i\bQ^{\text{new}}_1 \cdot \cX = i \Pi_{\LLh} (\bQ^{\text{old}} \cdot \cX)= -\Pi_{\LLh} (\omegab +\cX\cX-\sg^{-1} (i\bQ_1 \cdot A_\psi +A_\psi^2) \sg).
\end{align}
This remains a symmetry of the action since we only modified the variation of the $\cX$ and
\begin{align} \tr\{(\bQ_1^{\text{old}} \cdot \cX) \p \cX\} = \tr\{ (\bQ_1^{\text{old}} \cdot \cX) \Pi_{\LLh} \p\cX\}
= \tr\{ (\bQ_1^{\text{new}} \cdot \cX) \p \cX\}.\end{align}
This result holds for a general sub-algebra $\LLh \subset \Lg$.  When $\LLh$ is a Cartan subalgebra there are some important simplifications.  For instance, we can drop the $\cX\cX$ term from $i\bQ_1 \cdot \cX$; also $(\bQ_1)^2 \cdot \cX = \pb \cX$.  Note, however, that $(\bQ_1)^2 \cdot \sg$ is not just a standard translation; even for $A = 0$ and $\LLh$ Cartan, we find $(\bQ_1)^2 \cdot \sg = \sg \Pi_{\LLh} \sg^{-1} \pb \sg$.

\subsection{Enhanced supersymmetry}
We will now show that the supersymmetry can be further enhanced to the (0,2)+(0,4) structure.  The first step is to establish the necessary
$\GU(1)\times\SU(2)$ R-symmetries with generators $r$ and $R_a$ as in section~\ref{ss:0204}.  The $\GU(1)$ generator $r$ corresponds to a trace-compatible complex structure on $\LLh$~\cite{Berglund:1995dv}.  That is, a map $\cI : \LLh \to \LLh$ satisfying $\cI^2 = -1$ and
\begin{align}\tr_{\LLh} \{ x \cI(y)\} = -\tr_{\LLh} \{\cI(x) y\} \qquad \text{for all} \quad x,y \in \LLh.\end{align}
This is an integrable complex structure on the corresponding Lie group $H$ if $\cI$ satisfies an analogue of the vanishing of the Nijenhuis tensor.\footnote{A complex structure $\cI$ on a Lie algebra with generators $T^i$ and bracket
$\CO{T_i}{T_j} = C_{ij}^{~~k} T_k$ is integrable iff
$C_{ij}^{~~n} \cI^k_n \cI^i_m - C_{im}^{~~n} \cI^k_n \cI^i_j + C_{mj}^{~~k} - C_{in}^{~~k} \cI^i_m \cI^n_j = 0.$  Such structures exist on all even-dimensional Lie algebras, leading to many examples
of non-K\"ahler complex manifolds~\cite{Salamon:2002hg}.}  This holds for $\LLh$ abelian.  Having chosen such an $\cI$, we take the non-trivial action of the R-symmetry generators as
\begin{align} r \cdot \cX = -i \cI(\cX), \qquad R_a \cdot \psi^\mu = -i \cK^\mu_{a\nu} \psi^\nu,\end{align}
where the $\cK_a$ are the three anti-commuting complex structures of the base $M$.  Recall from section~\ref{ss:0204} that the three Hermitian forms $J_{a\mu\lambda} \equiv \cK_{a\mu}^\nu g_{\nu\lambda}$ satisfy $d J_a = \beta \wedge J_a$.
These are symmetries of the full fibered action provided that the curvature $F$ of the fibration is ASD, and $\cH = -\ast_{g} \beta$, as in section~\ref{ss:0204}.

\subsubsection*{Diagonal (0,2) supersymmetries}
The remaining supercharges can be constructed via commutators of the R-charges and $\bQ_1$, but  there is a slight complication as compared to the construction given above:  Because $\bP \equiv \bQ_1^2$ does not simply act as $\pb$ on $\sg$, it is not obvious that the R-symmetries commute with $\bP$.  However, an explicit computation shows this to be the case.  We just give the details for
\begin{align}
\CO{(i\bQ_1)^2}{r} \cdot \sg  &= i\bQ_1 \cdot (i\sg \cI(\cX)) + \CO{i\bQ_1}{r}\cdot (\sg \cX -A_\psi \sg) \nonumber\\
& = i (\sg \cX -A_\psi\sg) \cI(\cX) -\sg \cI(\bQ_1 \cdot \cX) + i \sg \cI(\cX) \cX + \sg \cI(\bQ_1\cdot \cX) + iA_\psi\sg \cI(\cX) \nonumber\\
& =0.
\end{align}
Using the ASD property of $F$ we can also show $\CO{(i\bQ_1)^2}{R_a} \cdot \sg =0$.

Let us show that $\bQ_1$, $\bP$, $\bR \equiv r + R_3$ and $\bQ_2 \equiv i \CO{\bQ_1}{\bR}$ satisfy a (0,2) algebra.  The statement is obvious on the base fields.  On the WZW fields we find
\begin{align}
\bQ_2 \cdot g & = - \bR\cdot (i\bQ_1 \cdot \sg) = i \sg \cI \cX - A_{\cK_3 \psi} \sg, \nonumber\\
\bQ_2 \cdot \cX & = i\bQ_1 \cdot (-i\cI (\cX)) - \bR \cdot (i\bQ_1 \cdot \cX) = \cI (\bQ_1 \cX).
\end{align}
Because $\bR$ commutes with $\bP$, the algebra will close as expected provided we can show $i\CO{\bR}{\bQ_2} = i\bQ_1$.  This indeed holds:
\begin{align}
i \CO{\bR}{\bQ_2} \cdot \sg &= i \bR \bQ_2\cdot \sg = i\bR [ i \sg \cI (\cX) - i A_{\cJ \psi} \sg] =
i\sg \cI^2 (\cX) -i A_{\cJ^2\psi} \sg \nonumber\\
&= -i \sg \cX + i A_\psi \sg = \bQ_1\cdot \sg; \nonumber\\
i \CO{\bR}{\bQ_2} \cdot \cX &= i\bR\cdot ( \cI (\bQ_1 \cdot \cX) ) - i \bQ_2\cdot (-i \cI( \cX)) = - \cI^2 (\bQ_1 \cdot\cX) = \bQ_1 \cdot\cX.
\end{align}
Clearly we generate a second (0,2) symmetry by sending $r \to -r$.

\subsubsection*{Further enhancement to (0,2)+(0,4)}
We will now demonstrate further enhancement with (0,2) generators
$q_A$, $r$, $p$ with non-trivial commutation relations
\begin{align}
\CO{r}{q_A} = i \ep_{AB} q_B, \quad \AC{q_A}{q_B} = 2\delta_{AB} p
\end{align}
and (0,4) generators $R_a$, $Q_0$, $Q_a$ and $P$ with non-trivial commutators
\begin{align}
\CO{R_a}{R_b} & = 2i \ep_{abc} R_c, \quad
\CO{R_a}{Q_0}  = iQ_a, \quad
\CO{R_a}{Q_b} = -i \delta_{ab} Q_0 + i \ep_{abc} Q_c, \nonumber\\
\AC{Q_a}{Q_b} &= 2 \delta_{ab} P, \quad Q_0^2 = P.
\end{align}
The strategy is the same as in~\cite{Melnikov:2010pq}.  Using the two diagonal (0,2) sub-algebras constructed above, we define 
the generators  
\begin{align}
\label{eq:defs}
q_2 &\equiv -i \CO{r}{\bQ_1}, &
q_1 &\equiv i  \CO{r}{\bQ_2},  & p &\equiv q_2^2,\nonumber\\
Q_a &\equiv -i \CO{R_a}{\bQ_1}, &
Q_0 &\equiv \bQ_1 - q_1, & P &\equiv \bP - p.
\end{align}
Since $r$ annihilates the base fields, we see that $r,q_A, p$ leave $(\phi,\psi)$ invariant, while $Q_0, Q_a, R_a$ and $P$ generate a (0,4) algebra on them, with the explicit generators acting as
\begin{align}
Q_0 \cdot \phi^\mu &= -i\psi^\mu, & Q_0 \cdot \psi^\mu &= i\pb\phi^\mu , \nonumber\\
Q_a \cdot \phi^\mu &= -i \cK_{a\nu}^\mu \psi^\nu, & Q_a \cdot \psi^\mu &= -i\cK^\mu_{a\nu}\pb\phi^\nu -i \cK^\mu_{a\nu,\rho}\psi^\nu\psi^\rho.
\end{align}
The action on the WZW fields is 
\begin{align}
q_1 \cdot \sg &= -i \sg\cX, & q_2 \cdot \sg &=+i \sg \cI (\cX), &
Q_0 \cdot \sg &= i A_\psi \sg, & Q_a \cdot \sg &=-i A_{\cK_a \psi}\sg,\nonumber\\
q_1 \cdot \cX &= \bQ_1 \cdot\cX, &q_2 \cdot \cX &= \cI (\bQ_1\cdot \cX),&
Q_0\cdot \cX &= 0, & Q_a \cdot \cX &= 0.
\end{align}
Using Jacobi identities we can show that the algebra will close to (0,2)+(0,4) if and only if
\begin{align}
\label{eq:A2A4req}
\CO{r}{R_a} & = 0, \qquad \CO{R_a}{R_b} = 2i\ep_{abc} R_c,
 \nonumber\\
\CO{R_a}{q_A} &= 0, \qquad \CO{r}{q_1} = i q_2, \qquad
\CO{R_a}{Q_b}  + \CO{R_b}{Q_a}=0, \quad a\neq b.
\end{align}
These are satisfied on $(\phi,\psi)$, so all that remains is to check the relations 
on $\sg$ and $\cX$.  The first two are obviously satisfied; next we have
\begin{align}
\CO{R_a}{q_A}\cdot \sg= R_a \cdot(q_A \cdot \sg) = 0; \qquad
\CO{R_a}{q_A} \cdot \cX = R_a\cdot (q_A\cdot\cX) = 0.
\end{align}
It is also easy to see
\begin{align}
\CO{r}{q_1} \cdot \sg &  = r\cdot(q_1 \cdot\sg) = -ir\cdot(\sg \cX) =-\sg \cI (\cX) = i q_2\cdot\sg,\nonumber\\
\CO{r}{q_1} \cdot \cX & = -q_1\cdot(r\cdot\cX) = i \cI(\bQ_1\cdot\cX) = i q_2 \cdot \cX.
\end{align}
Finally, we have $\CO{R_a}{Q_b} \cdot \cX = 0$ and 
\begin{align}
\CO{R_a}{Q_b} \cdot \sg = R_a \cdot (Q_b\cdot\sg) = - A_{\cK_b \cK_a \psi}.
\end{align}
Since $\AC{\cK_a}{\cK_b} = - 2 \delta_{ab}$, we see that for $a\neq b$
\begin{align}
(\CO{R_a}{Q_b} + \CO{R_b}{Q_a} )\cdot \sg = -A_{\AC{\cK_b}{\cK_a} \psi }= 0.
\end{align}
Thus, the fibered WZW construction of the $N_V = 2$, $N_H =129$ example from~\cite{Kachru:1995wm} realizes the expected (0,2)+(0,4) supersymmetry.  It is indeed a heterotic flux vacuum, where the symmetry currents of $T^2$ (in this case enhanced to $\su(2)\oplus\Lu(1)$) are gauged over a $\KT$ base.  The Chern-Simons form for the connection associated to the fibered $\su(2)$ contributes to $\cH$ in the same fashion as that of the connection $\cA$ for the left-moving fermions, but the requisite shift and accompanying terms in the action can already be seen at tree-level in $\alpha'$.

\section{Discussion}
We explored a number of aspects of perturbative heterotic vacua with N=2 spacetime supersymmetry in four dimensions.  The requirement of  (0,2)+(0,4) worldsheet supersymmetry leads to stringent constraints on the background geometry and bundle, essentially reducing the non-trivial geometric structure to a choice of bundle over a K3 surface.  The existence of these vacua requires a balancing between tree-level and one-loop terms in the $\alpha'$ expansion, and the massless deformations are constrained by flux quantization.  We explored these effects from the worldsheet perspective, and the qualitative conclusion is that, as far as geometric vacua are concerned, we have a fairly complete description.  This is should be contrasted with N=1 heterotic vacua, where there is not even a topological classification of base manifolds; moreover, genuine non-geometric vacua are expected to be at least as ubiquitous as geometric ones~\cite{McOrist:2010jw}.

The main motivation for our study was to understand how heterotic flux vacua fit into type II/heretoric duality.  Fairly basic considerations lead to the hypothesis that the type II duals of heterotic vacua should be based on K3-fibered three-folds lacking a compatible elliptic fibration with section.  Following this, we constructed a number of interesting potential dual pairs.  It will be interesting to test the proposal in more detail and use it to extend the class of known dual pairs. One of the surprises of our exploration was the possibility of non-sequential Higgsing raised in section~\ref{ss:twovectors}; it would be nice to settle this either affirmatively or negatively.  

Another interesting direction to pursue is to explore the duality by starting with the $d=8$ equivalence between
F-theory on a K3 and the heterotic string on $T^2$.\footnote{This has already been used in explorations of N=2, $d=4$ dualities~\cite{Sethi:1996es}, and more recently for the purpose of identifying non-geometric heterotic backgrounds in~\cite{McOrist:2010jw}.}  Fibering these dual descriptions over a base K3 should provide a concrete proposal not only for potential dual pairs but also for the map of the corresponding moduli spaces.  This set of examples may also be a useful laboratory for exploring F-theoretic G-flux in a controlled (i.e. N=2 ) setting, as in, e.g.~\cite{Aspinwall:2005qw}.  
We hope to return to these questions in the future.  

\appendix
\section{Details of the background field expansion} \label{app:background}
The computation of the effective action quoted in~(\ref{eq:Snloc}) proceeds in three steps,
all of them reasonably well-understood.

First, we split the fields into a background and 
quantum contributions, using geodesic normal coordinates.  We then expand the action
about a background that satisfies the classical equations of motion, keeping terms quadratic
in the quantum fields.  This is sufficient to compute the effective action to quadratic order in
$\cA$ and $\cS^+$.  The necessary methodology is well described in~\cite{Ketov:2000qn}.

Second, we evaluate the quadratic contributions to the effective action.  As these are
one-loop computations, there is no need for supergraph machinery; instead,
we compute directly using superspace OPEs, taking care to regularize divergences 
and evaluating contributions from certain canonical contact terms.  The
latter were described in~\cite{Green:1987qu}.

Finally, by using the background equations of motion, we isolate the non-covariant terms.
We then check that the gauge variation of these terms can be canceled by adding a local
counter-term and shifting $B$ appropriately.  Our final result agrees with~\cite{Hull:1986xn},
but we hope that presenting the additional details makes the derivation a bit clearer.

\subsection{Covariant background superfields}
Let $\Phit(s)$ and $\Lambdat(s)$ denote a one-parameter family of fields with derivatives
\begin{align}
\Sigma_s \equiv \frac{d}{ds} \Phit(s), \qquad \cX_s \equiv \frac{d}{ds} \Lambdat + \Sigma^\mu \cA_\mu(\Phit) \Lambdat
\end{align}
that satisfy the parallel transport equations
\begin{align}
\dot \Sigma_s^\lambda + \Gamma^\lambda_{\mu\nu}(\Phit) \Sigma^\mu_s \Sigma^\nu_s = 0, \qquad
\nabla_s \cX_s = 0, 
\end{align}
with $\nabla_s$ the covariant derivative constructed with the gauge connection $\cA(\Phit)$.
The background $(\Phi,\Lambda)$ specifies the initial 
values $\Phit(0) = \Phi$ and $\Lambdat(0) = \Lambda$, and we take the quantum fields to be
$\Sigma \equiv \Sigma_{s=0}$ and $\cX \equiv \cX_{s=0}$.  With this in mind, we obtain the
action for the fluctuations by solving the geodesic equations in a power-series around $s=0$ and
expanding 
\begin{align}S(\Phit,\Lambdat) = \sum_{n=0}^\infty \frac{s^n}{n!} S_n(\Phi,\Lambda;\Sigma,\cX).\end{align}
The $n$-th term is the $O(n)$ term in the expansion of the action in the fluctuating
fields.  The great virtue of this ``geodesic expansion'', appreciated early on~\cite{Friedan:1980jm,AlvarezGaume:1981hn}, is that the resulting quantum action is explicitly target space diffeomorphism-invariant.
As emphasized in~\cite{Ketov:2000qn}, extracting the terms order by order is greatly simplified by using a covariant derivative and not the naive $d/ ds$.  
If we assume that the background fields satisfy the classical equations of motion~(\ref{eq:ceom}), then
the $O(s)$ terms vanish, and the leading terms in the expansion of~(\ref{eq:eucahsuper}) have the action
$S_2 = \ff{1}{4\pi} \int d^2 z \cL_2$ with 
\begin{align}
\cL_2 & = g_{\alpha\beta} D_z^-\Sigma^\alpha D_\theta^+ \Sigma^\beta +\Sigma^\alpha\Sigma^\beta \p\Phi^\mu \cD\Phi^\nu \left[ R_{\mu\alpha\beta\nu} + \frac{1}{2} \nabla_\alpha H_{\beta\mu\nu} +\frac{1}{4} H_{\gamma\mu\alpha} H_{\delta\nu\beta} g^{\gamma\delta} \right] \nonumber\\
& \qquad 
- \cX^T D_\theta \cX 
+ 2 \cD \Phi^\mu \Sigma^\nu \cX^T \cF_{\nu\mu}\Lambda
+\ff{1}{2} \Sigma^\nu D_\theta\Sigma^\mu \Lambda^T \cF_{\nu\mu} \Lambda
+ \ff{1}{2}\cD\Phi^\mu \Sigma^\nu\Sigma^\lambda \Lambda^T \nabla_\lambda\cF_{\nu\mu} \Lambda,
\end{align}
where 
\begin{align}
D_\theta \cX = \cD \cX + \cD \Phi \cA_\mu \cX,
\end{align}
$H \equiv dB$, and 
\begin{align}
D_z^- \Sigma^\alpha = \p \Sigma^\alpha +\p\Phi^\mu (\Gamma^\alpha_{\mu \gamma} -\ff{1}{2} H^\alpha_{~\mu\gamma}) \Sigma^\gamma, \qquad
D_\theta^+ \Sigma^\beta = \cD\Sigma^\beta + \cD\Phi^\nu (\Gamma^\beta_{\nu\delta} + \ff{1}{2} H^\beta_{~\nu\delta}) \Sigma^\delta.
\end{align}
The final step is to re-express the $\Sigma^\mu$ in terms of the more convenient frame bundle fields $\Sigma^a$.  We introduce a vielbein $e^a_\mu$ and its inverse $E^{a\mu}$ such that $g_{\mu\nu} = e^a_\mu e^a_\nu$ and write the action in terms of $\Sigma^a = e^a_\mu \Sigma^\mu$.  The result is
\begin{align}
\label{eq:action2}
\cL_2 & = (\p \Sigma^a + \p\Phi^\lambda \cS^{-ab}_\lambda \Sigma^b)(\cD\Sigma^a + \cD\Phi^\mu \cS^{+ac}_\mu \Sigma^c)
+ \Sigma^a \Sigma^b \p\Phi^\mu \cD \Phi^\nu R^+_{\mu ab \nu} \nonumber\\[0.2cm]
& \quad - \cX^T( \cD \cX + \cD \Phi^\lambda \cA_\lambda\cX)
+ 2 \cD\Phi^\mu \Sigma^a \cX^T \cF_{a\mu} \Lambda \nonumber\\[0.2cm]
& \quad
+ \ff{1}{2} \Sigma^a (\cD\Sigma^b + \cD\Phi^\lambda\omega^{bc}_\lambda \Sigma^c) \Lambda^T \cF_{ab} \Lambda
+ \ff{1}{2} \cD\Phi^\mu \Sigma^a \Sigma^b \Lambda^T \nabla_b\cF_{a\mu} \Lambda,
\end{align}
where
\begin{align}
R^+_{\mu a b \nu} = E^\alpha_a E^\beta_b \left[ R_{\mu\alpha\beta\nu} + \frac{1}{2} \nabla_\alpha H_{\beta\mu\nu} +\frac{1}{4} H_{\gamma\mu\alpha} H_{\delta\nu\beta} g^{\gamma\delta} \right],
\end{align}
$\omega$ is the torsion-free, metric compatible spin connection, and, as in~(\ref{eq:Spm}),
\begin{align}
\cS^{\pm ab}_\lambda = \omega^{ab}_\lambda \pm \ff{1}{2} E^{a\sigma} E^{b\nu} H_{\sigma\lambda\nu}.\end{align}

\subsection{The quadratic effective action}
Having written down the quadratic action, we are ready to compute the one-loop corrections to the effective
action that are quadratic in the background fields $\cA$, $\cS^\pm$ and $\omega$.  This is a very special set of
terms because we can compute them just by considering the terms in $\cL_2$; we do not need the $O(s^3)$ or
higher terms in the quantum action.  

\subsubsection*{Free theory and supersymmetric contact terms}
We expand around the free theory with action
\begin{align}
S_{\tfree} = \frac{1}{4\pi} \int d^2 zd\theta \left[ \p\Sigma^a \cD \Sigma^a - \cX^T \cD \cX\right].
\end{align}
The super OPEs 
\begin{align}
\Sigma^a(\bz_1) \Sigma^b(\bz_2) \sim - \delta^{ab} \log (z_{12} (\zb_{12} -\theta_1\theta_2)) \qquad
\cX^A (\bz_1) \cX^B(\bz_2) \sim \frac{\delta^{AB}}{z_{12}}
\end{align}
determine all correlators by Wick's theorem.  It is a familiar fact that 
sufficiently singular functions of $z_{12}$ are non-holomorphic due to contact terms (e.g.
$\pb_1 z_{12}^{-1} = 2\pi \delta^2(z_{12},\zb_{12})$); similarly, they also carry a $\theta$ dependence
if we wish them to be supersymmetric~\cite{Green:1987qu}.  That is
\begin{align}
\xi (\cQ_1 + \cQ_2) \frac{1}{z_{12}} = 0 \implies 
\p_{\theta_1} \frac{1}{z_{12}} = -2\pi \theta_2 \delta^2(z_{12},\zb_{12}).
\end{align}
In fact, one can define a $\theta$-independent ``principal part'' of $z_{12}^{-1}$ by
\begin{align}
\frac{1}{z_{12}} = \Ppart \frac{1}{z_{12}} - 2\pi \theta_1\theta_2 \delta^2(z_{12},\zb_{12}).
\end{align}
An important consequence for what follows is 
\begin{align}\cD_1 z_{12}^{-1} = 2\pi (\theta_1-\theta_2) \delta^2(z_{12},\zb_{12}).\end{align}

\subsubsection*{The interaction Lagrangian}
To express the interaction Lagrangian of~(\ref{eq:action2}) succinctly, we introduce a short-hand
for various pull-backs from the target space;  for example, 
$\cS^{\pm ab}_\theta \equiv \cD\Phi^\mu \cS^{\pm ab}_\mu$, $\cS^{\pm ab}_z \equiv \p \Phi^\mu \cS^{\pm ab}_\mu$, etc.  With this notation the interaction terms linear in the background are
\begin{align}
\cL_{\tint} &= \p \Sigma^a \cS^{+ab}_\theta \Sigma^b  + \cD\Sigma^a  ( \cS^{-ab}_z -\ff{1}{2}\Lambda^T \cF_{ab} \Lambda)\Sigma^b
-\cX^T \cA_\theta \cX \nonumber\\
&\quad-2 \Sigma^a \cX^T \cF_{a\theta} \Lambda 
+\Sigma^a\Sigma^b (R^+_{z(ab)\theta} - \ff{1}{2}\Lambda^T \nabla_{(b} \cF_{a)\theta} \Lambda ).
\end{align}
At quadratic order, the terms in the first line have no non-trivial contractions with those in the second 
line.\footnote{Either a full contraction is impossible, or it is zero due to symmetry properties  under $a\leftrightarrow b$.}
Since the contractions among terms from the second line yield explicitly covariant
terms, we can concentrate on the quadratic terms due to
\begin{align}
\cL'_{\tint} & =\p \Sigma^a \cS^{+ab}_\theta \Sigma^b  + \cD\Sigma^a  \cT^{ab} \Sigma^b
-\cX^T \cA_\theta \cX , \qquad \cT^{ab} \equiv \cS^{-ab}_z -\ff{1}{2} \Lambda^T \cF_{ab} \Lambda.
\end{align}
At quadratic order in the background, the possible contractions of these interactions yield either 
$O(\cA^2)$ or $O(\cS_+^2)$ terms; we consider these in turn.

\subsubsection*{The $\cX$ contributions}
The $O(\cA^2)$ correction to the partition function is
\begin{align}
\Delta Z_{\cX} = \frac{1}{2} \int \frac{d^2 z_1  d^2 z_2 d\theta_2d\theta_1}{(4\pi)^2} 
\la \cX^T_1 \cA_{1\theta} \cX_1 \times \cX^T_2 \cA_{2\theta} \cX_2 \ra,
\end{align}
where the correlator is to be evaluated with free field OPEs.  The result, interpreted as a term in the
effective action, is
\begin{align}
\Delta S_\cX = - \int \frac{d^2 z_1  d^2 z_2 d\theta_2d\theta_1}{(4\pi)^2} \frac{\tr\{\cA_{1\theta}\cA_{2\theta}\}}{z_{12}^2}.
\end{align}
As in the main text, the subscripts $1$ and $2$ refer to the superspace insertion of the field;  thus $\cA_{1\theta} \equiv \cA_\mu(\Phi(\bz_1)) \cD_1 \Phi(\bz_1)$, and $\cD_1 = \p_{\theta^1} + \theta^1 \pb_{1}$.

\subsubsection*{The $\Sigma$ contributions}
The $O(\cS_+^2)$ terms are somewhat more involved.  The main complication is due to the logarithm in
the $\Sigma_1 \Sigma_2 $ OPE.  The resulting logarithms lead to IR divergences in the $\bz_{1,2}$ integrals.
To handle these we regulate the OPE in a supersymmetric manner.  Introducing the supersymmetric invariants
$\theta_{12} \equiv \theta_1 -\theta_2$ and $\zeta_{12} \equiv \zb_{12}-\theta_1\theta_2$, we take the
regulated two-point function to be
\begin{align}
\la \Sigma^a_1\Sigma^b_2 \ra = -\delta^{ab} \Delta_{12} , \qquad \Delta_{12} \equiv \log( z_{12} \zetab_{12} + \ell^2 ),
\end{align}
where $\ell$ is a regulating lengthscale.  Note that this is still explicitly supersymmetric, because 
\begin{align}
R \equiv z_{12}\zetab_{12} + \ell^2
\end{align}
is annihilated by $(\cQ_1 + \cQ_2)$.
With this regulator, we obtain
\begin{align}
\Delta S_\Sigma & = \int \frac{d^2 z_1  d^2 z_2 d\theta_2d\theta_1}{(4\pi)^2} \left[ 
 \ff{1}{2}\tr\{\cS^+_{1\theta}\cS^+_{2\theta}\} X
+\tr\{\cS^+_{1\theta} \cT_{2}\} Y
+\ff{1}{2} \tr\{\cT_1\cT_2\} Z\right],
\end{align}
where
\begin{align}
 X &= \frac{1}{2}  \p_1\p_2 \Delta_{12}^2 - 2 \p_1 \Delta_{12} \p_2 \Delta_{12},\nonumber\\
 Y &=  -\frac{1}{2} \p_1 \cD_2 \Delta_{12}^2 + 2 \Delta_{12} \p_1 \cD_2 \Delta_{12},\nonumber\\
 Z &= \Delta_{12} \cD_1 \cD_2 \Delta_{12} - \cD_1 \Delta_{12} \cD_2 \Delta_{12}.
\end{align}
To simplify these terms, we first note that since $\cD_1 \Delta_{12} = z_{12}\theta_{12} R^{-1}$, the
second term in $Z$ vanishes.  The second term in $X$ has a simple $\ell \to 0$ 
limit:
\begin{align}
-2\p_1 \Delta_{12} \p_2 \Delta_{12} &=  \frac{2 \zetab_{12}^2}{(z_{12} \zetab_{12} +\ell^2)^2} \underset{\ell \to 0}{~~\longrightarrow~~}  \frac{2}{z_{12}^2};
\end{align}
while the second term in $Y$ is actually a UV-divergent local term since
\begin{align}
\p_1 \cD_2 \Delta_{12} = \theta_{12} \frac{\ell^2}{(z_{12}\zb_{12} + \ell^2)^2} \underset{\ell \to 0}{~~\longrightarrow~~}
2\pi \theta_{12} \delta^2(z_{12}).
\end{align}
Thus, up to a local counter-term, we find  $\Delta S_{\Sigma} = \Delta S_1 + \Delta S_2$ with
\begin{align}
\Delta S_1 & =  \int \frac{d^2 z_1  d^2 z_2 d\theta_2d\theta_1}{(4\pi z_{12})^2}\tr\{\cS^+_{1\theta} \cS^+_{2\theta} \}, \nonumber\\
\Delta S_2 & =  \int \frac{d^2 z_1  d^2 z_2 d\theta_2d\theta_1}{4(4\pi)^2}
 \tr\{(\p_1\cS^+_{1\theta}-\cD_1\cT_1) ( \p_2 \cS^+_{2\theta} - \cD_2 \cT_2) \} \Delta_{12}^2.
\end{align}
The second contribution looks complicated, but fortunately we need not consider it.  Up to terms of
higher order in the background and using the classical equations of motion for $\Phi$ and $\Lambda$,
we find
\begin{align}
\p S^{+ab}_\theta - \cD T^{ab}  = \cD\Phi^\mu \p \Phi^\lambda( d\omega^{ab}_{\lambda\mu} + \ff{1}{2} H^{a~b}_{~\mu~,\lambda} + \ff{1}{2} H^{a~b}_{~\lambda~,\mu}).
\end{align}
This is invariant under the linearized Lorentz transformations, and we expect that incorporation of
higher order terms in the background will provide a fully covariant form for $\Delta S_2$.   So, the non-covariant terms in the $O(\cS_+^2)$ contribution to the one-loop effective action have, up to a crucial minus sign, the same form as $\Delta S_{\cA}$, and the combined non-covariant terms are
\begin{align}
\Delta S & =  \int \frac{d^2 z_1  d^2 z_2 d\theta_2d\theta_1}{(4\pi)^2} \frac{\tr\{\cS^+_{1\theta} \cS^+_{2\theta} \}-\tr\{\cA_{1\theta}\cA_{2\theta}\}}{z_{12}^2}.
\end{align}
To obtain the final form quoted in the text, we use $z_{12}^{-2} = \p_2 z_{12}^{-1}$ and rewrite $\p \cA_\theta$ in a more convenient way up to background fields' equations of motion and higher order terms in $\cA$:
\begin{align}
\p\cA_\theta = \p \cD\Phi^\lambda \cA_\lambda = \cD\Phi^\lambda \p\Phi^\rho \cA_{\lambda,\rho} =
\cD\Phi^\lambda \p\Phi^\rho d\cA_{\rho\lambda} + \cD (\cA_{z}).\end{align}
This agrees with the results originally obtained in~\cite{Hull:1986xn} and quoted above in~(\ref{eq:Snloc}).

\section{N=2 Higgsing, sequential and otherwise} \label{app:higgs}
Consider an N=2 four-dimensional gauge theory with gauge group $G$ (Lie algebra $\Lg$) and hypermultiplets transforming in $\oplus_\alpha \rep{r}_\alpha$, where $\rep{r}_\alpha$ label irreducible representations of $\Lg$.  Each hypermultiplet has four real scalars, and each vector multiplet contributes an additional complex scalar.  $N=2$ supersymmetric vacua correspond to zeroes of the scalar potential, and the Higgs branch is the set of vacua where the vector multiplet scalars are set to zero.  

To describe the remaining constraints on the hypermultiplet expectation values on the Higgs branch, it is convenient to use an N=1 superspace description, where a hypermultiplet in $\rep{r}$ is represented by two chiral multiplets $Q$ and $\Qt$ transforming in $\rep{r}$ and $\brep{r}$ respectively.\footnote{This is all well-known; a clear presentation is given in~\cite{Seiberg:1994rs,Seiberg:1994aj}.  We find it convenient to think of $Q$ as column and $\Qt$ as row vectors; we will also label the expectation values of the scalar fields by the same letters as the chiral multiplets.} 
The constraints on the scalar expectation values then arise as N=1 $D$ and $F$ terms~\cite{Argyres:1996eh}.  Denoting the Hermitian generators of $\Lg$ in $\rep{r}_\alpha$ by $M_{\rep{r}_\alpha}$, the supersymmetry conditions are that for every $M_{\rep{r}_\alpha}$  we have
\begin{align}
\text{(F-terms) }\quad \sum_\alpha \Qt_\alpha  M_{\rep{r}_\alpha} Q_\alpha  = 0, \qquad
\text{(D-terms)} \quad \sum_\alpha Q^\dag_\alpha M_{\rep{r}_\alpha} Q_\alpha - \Qt_\alpha M_{\rep{r}_\alpha} \Qt^\dag_\alpha = 0.
\end{align}
For general $G$ and matter content this describes a complicated hyper-K\"ahler quotient space.  In general this is a reducible affine variety with many components of different dimensions and with different unbroken gauge symmetry.  Some well-studied cases are the classical gauge groups with matter in fundamental representations~\cite{Argyres:1996eh,Argyres:1996hc}; more recently there has been interesting work on more exotic theories, e.g.~\cite{Benini:2009gi,Hanany:2010qu,Chacaltana:2012zy}.  However, we are not aware of any algorithmic answer even to the very coarse question of when $G$ can be broken completely.  

Since the $N=2$ Higgs mechanism requires a vector multiplet to eat a full hypermultiplet, it is clear that a necessary condition is that the number of $G$-charged hypers should be greater than $\dim G$.  However, this is certainly not sufficient.  For instance~\cite{Argyres:1996hc}, for $G = \SO(n_c)$ with $n_f$ hypermultiplets in $\rep{n_c}$ this necessary condition for complete Higgsing is $2n_f \ge n_c-1$, but full Higgsing is only possible when $n_f \ge n_c$.

It is much simpler to give sufficient conditions for partial Higgsing.  For instance, suppose we have a hypermultiplet in a real representation $\rep{r}$, so that the generators $M_{\rep{r}}$ can be taken to be pure imaginary and hence anti-symmetric.  Then it is easy to see that $Q = \Qt = v$ for any real vector $v \in \rep{r}$ will solve the F- and D-terms.  The unbroken gauge group is then the stabilizer subgroup $H\subset G$ of the real vector $v$.  In particular, we can always Higgs $\SO(n_c)$ with $n_f$ fundamental hypermultiplets to $H= \SO(n_c-1)$, $n_f-1$ fundamental and $n_f$ $H$-neutral hypermultiplets.

When $\rep{r}$ is complex or pseudo-real it is not in general possible to Higgs the theory by just giving an expectation value to a single hypermultiplet.  The classic example of this is $G = \SU(n_c)$ with a single hypermultiplet in the fundamental~\cite{Argyres:1996eh}.  Denoting the color index by $i$, the D- and F-term equations are equivalent to
\begin{align}
\Qt^i Q_j = \nu \delta^i_j, \qquad Q^{\dag i} Q_j - \Qt^i Q^\dag_j = \rho \delta^i_j, \qquad \nu \in \C,~~\rho\in \R.
\end{align}
Without loss of generality we can assume $Q \neq 0$; the first equation then requires $\Qt = 0$ and $\nu = 0$, in which case the second equation has no solution.

We can do better when there are two or more hypermultiplets transforming in $\rep{r}$.  Denoting the $N=1$ components of two of these by $(Q,\Qt)$ and $(q,\qt)$, we can solve the D-terms by setting $\Qt = 0$, $q = 0$, and $\qt^{\dag} = Q = v$ for some $v \in \rep{r}$. 

\subsubsection*{The $\GE_7$ theory with $k$ half-hypermultiplets in $\rep{56}$}
Having covered those basic generalities, we turn to the $\GE_7$ example discussed in the text.
For $k \ge 4$ there are at least two full hypermultiplets in the pseudo-real $\rep{56}$, and by the
discussion above we see that we can Higgs $\GE_7$ to a stabilizer of a complex vector $v \in \rep{56}$.
From the decomposition of $\rep{56} = \rep{27}+\brep{27} + 2\times\rep{1}$ under an $\GE_6$ subgroup,
we see that we can choose $v$ so that the stabilizer is $\GE_6$.  On this Higgs branch we obtain
$k-2$ hypermultiplets in $\rep{27}$ and $(k-1)$ $\GE_6$-singlets.  If we assume $k>4$, then using the
steps outlined above, we proceed to further sequential breaking via 
\begin{align}
\GE_6 \to \SO(10)\to \SO(9)\to\SO(8)  \to \SO(7) \to \GG_2 \to \SU(3)
\end{align}
with a matter spectrum in the final step given by
\begin{align}
6(k-5) \times\rep{3} + 5(2k-7)\times\rep{1}.
\end{align}
When $k > 5$ there is plenty of matter to break $\SU(3)$ completely, but for $k=5$ this sequence
does not allow full breaking.  When $k=4$ this chain terminates at $\SO(8)$.

\subsection*{Possible non-sequential Higgsing}
There is, however, another possibility:  instead of breaking the gauge groups in steps, we might try to contrive the expectation values in such a way as to break the full group at once.  In making such an attempt, there are two questions to consider:  can we assign expectation values so that the stabilizer (i.e. the little group) of the configuration is trivial?  can we do so while preserving supersymmetry?

As far as trivial stabilizer is concerned, the answer is affirmative.  A complex vector $v$ in the $\rep{27}$ of $\GE_6$ has four $\GE_6$ invariants that can be constructed from the invariant tensors $\delta^a_b$ and $d^{abc}$ of the fundamental representation:
\begin{align}
v_a v_b v_c d^{abc} \in \C, \qquad\text{and}\quad
v_a \vb^a, \quad
 v_a v_b d^{abc}  \vb^d \vb^e d_{dec} \in \R.
\end{align}
These can be identified in a reasonably straightforward fashion by decomposing $\rep{27}$ with respect to $\SO(10)$~\cite{Kugo:1994qr} or to $\SU(3)^3$~\cite{Kephart:1981gf}.  An octonionic discussion in terms of $\SL(3,O)$ representations was given in~\cite{Gursey:1980}.  The stabilizer of  $v$ depends on the values of the invariants.  It is certainly possible to choose them so that $v$ is stabilized by either $\GF_4$ or $\SO(10)$.  However, the most generic choice leads to a smaller stabilizer of $\SO(8)$.  Two more independent vectors of $\rep{27}$ are sufficient to reduce the stabilizer from $\SO(8)$ to $1$.

The real question, however, is whether the expectation values of the $3$ $\rep{27}$s can be chosen to lead to trivial stabilizer and to satisfy the supersymmetry conditions.  We have not been able to find such a solution, nor have we been able to show that complete breaking is impossible.

  The failure of the particular sequence of Higgsing above should not dismay us.\footnote{This non-pessimistic note was made in~\cite{Duff:1996rs}.}  For instance, in a SQCD theory with $G  = \SU(2r)$ and $n_f = 2r$ flavors there is a Higgs branch with an unbroken $\SU(r)$ symmetry and no charged matter, but there is also a branch where $G$ is completely broken~\cite{Argyres:1996eh}.  However, it may be~\cite{Duff:1996rs} that full Higgsing only takes place at some special locus in the moduli space where an enhanced gauge group is accompanied by appearance of extra hypermultiplets.  
  
\acknowledgments  It is a pleasure to thank  A.~Adams, L.~Anderson, P.~Aspinwall, A.~Degeratu, J.~Gray, D.~Israel, S.~Katz, A.~Kleinschmidt, V.~Kumar, J.~Lapan, D.~Morrison, T.~Nutma, E.~Sharpe,  W.~Taylor, and O.~Varela for useful discussions.  We would like to especially thank G.~Bossard for an extensive correspondence on the mysteries of $\GE_6$ orbits.  IVM would like to thank the Simons Center, KITP, and the University of Heidelberg for hospitality while some of this work was being completed.  RM thanks the Alexander von Humboldt foundation for support.  ST thanks T.~Weigand and A.~Hebecker for discussions and hospitality during his stay at the University of Heidelberg;  he also acknowledges the Klaus Tschira Foundation for general support during his stay.


\begin{thebibliography}{10}

\bibitem{Kachru:1995wm}
S.~Kachru and C.~Vafa, ``{Exact results for N=2 compactifications of heterotic
  strings},'' \href{http://dx.doi.org/10.1016/0550-3213(95)00307-E}{{\em
  Nucl.Phys.} {\bf B450} (1995)  69--89},
  \href{http://arxiv.org/abs/hep-th/9505105}{{\tt arXiv:hep-th/9505105
  [hep-th]}}.

\bibitem{Ferrara:1995yx}
S.~Ferrara, J.~A. Harvey, A.~Strominger, and C.~Vafa, ``{Second quantized
  mirror symmetry},''
  \href{http://dx.doi.org/10.1016/0370-2693(95)01074-Z}{{\em Phys.Lett.} {\bf
  B361} (1995)  59--65}, \href{http://arxiv.org/abs/hep-th/9505162}{{\tt
  arXiv:hep-th/9505162 [hep-th]}}.

\bibitem{Aspinwall:1996mn}
P.~S. Aspinwall, ``{K3 surfaces and string duality},''
\href{http://arxiv.org/abs/hep-th/9611137}{{\tt arXiv:hep-th/9611137}}.

\bibitem{Aspinwall:2000fd}
P.~S. Aspinwall, ``{Compactification, geometry and duality: N = 2},''
\href{http://arxiv.org/abs/hep-th/0001001}{{\tt arXiv:hep-th/0001001}}.

\bibitem{Klemm:1995tj}
A.~Klemm, W.~Lerche, and P.~Mayr, ``{K3 Fibrations and heterotic type II string
  duality},'' \href{http://dx.doi.org/10.1016/0370-2693(95)00937-G}{{\em
  Phys.Lett.} {\bf B357} (1995)  313--322},
  \href{http://arxiv.org/abs/hep-th/9506112}{{\tt arXiv:hep-th/9506112
  [hep-th]}}.

\bibitem{Aspinwall:1995vk}
P.~S. Aspinwall and J.~Louis, ``{On the ubiquity of K3 fibrations in string
  duality},'' \href{http://dx.doi.org/10.1016/0370-2693(95)01541-8}{{\em
  Phys.Lett.} {\bf B369} (1996)  233--242},
  \href{http://arxiv.org/abs/hep-th/9510234}{{\tt arXiv:hep-th/9510234
  [hep-th]}}.

\bibitem{Hull:1985jv}
C.~M. Hull and E.~Witten, ``{Supersymmetric sigma models and the heterotic
  string},'' {\em Phys. Lett.} {\bf B160} (1985)  398--402.

\bibitem{Sen:1986mg}
A.~Sen, ``{(2, 0) supersymmetry and space-time supersymmetry in the heterotic
  string theory},'' {\em Nucl. Phys.} {\bf B278} (1986)  289.

\bibitem{Banks:1987cy}
T.~Banks, L.~J. Dixon, D.~Friedan, and E.~J. Martinec, ``{Phenomenology and
  conformal field theory or can string theory predict the weak mixing
  angle?},''
{\em Nucl. Phys.} {\bf B299} (1988)  613--626.

\bibitem{Banks:1988yz}
T.~Banks and L.~J. Dixon, ``{Constraints on string vacua with space-time
  supersymmetry},'' {\em Nucl. Phys.} {\bf B307} (1988)  93--108.

\bibitem{Lauer:1988aw}
J.~Lauer, D.~Lust, and S.~Theisen, ``Supersymmetric string theories,
  superconformal algebras and exceptional groups,''
  \href{http://dx.doi.org/10.1016/0550-3213(88)90340-9}{{\em Nucl.Phys.} {\bf
  B309} (1988)  771}.

\bibitem{Witten:1995gx}
E.~Witten, ``{Small instantons in string theory},''
  \href{http://dx.doi.org/10.1016/0550-3213(95)00625-7}{{\em Nucl.Phys.} {\bf
  B460} (1996)  541--559},
\href{http://arxiv.org/abs/hep-th/9511030}{{\tt arXiv:hep-th/9511030
  [hep-th]}}.

\bibitem{Aspinwall:1997ye}
P.~S. Aspinwall and D.~R. Morrison, ``{Point - like instantons on K3
  orbifolds},'' \href{http://dx.doi.org/10.1016/S0550-3213(97)00516-6}{{\em
  Nucl.Phys.} {\bf B503} (1997)  533--564},
  \href{http://arxiv.org/abs/hep-th/9705104}{{\tt arXiv:hep-th/9705104
  [hep-th]}}.

\bibitem{Friedan:1985ge}
D.~Friedan, E.~J. Martinec, and S.~H. Shenker, ``{Conformal invariance,
  supersymmetry and string theory},'' {\em Nucl. Phys.} {\bf B271} (1986)  93.

\bibitem{Polchinski:1998rr}
J.~Polchinski, {\em String Theory}, vol.~2.
\newblock Cambridge University Press, Cambridge, UK, 1998.

\bibitem{Dixon:1987yp}
L.~J. Dixon, V.~Kaplunovsky, and C.~Vafa, ``{On four-dimensional gauge theories
  from type II superstrings},'' {\em Nucl.Phys.} {\bf B294} (1987)  43--82.

\bibitem{deWit:1995zg}
B.~de~Wit, V.~Kaplunovsky, J.~Louis, and D.~Lust, ``{Perturbative couplings of
  vector multiplets in N=2 heterotic string vacua},''
  \href{http://dx.doi.org/10.1016/0550-3213(95)00291-Y}{{\em Nucl.Phys.} {\bf
  B451} (1995)  53--95}, \href{http://arxiv.org/abs/hep-th/9504006}{{\tt
  arXiv:hep-th/9504006 [hep-th]}}.

\bibitem{Distler:2007av}
J.~Distler and E.~Sharpe, ``{Heterotic compactifications with principal bundles
  for general groups and general levels},'' {\em Adv.Theor.Math.Phys.} {\bf 14}
  (2010)  335--398, \href{http://arxiv.org/abs/hep-th/0701244}{{\tt
  arXiv:hep-th/0701244 [hep-th]}}.

\bibitem{Hull:1986xn}
C.~M. Hull and P.~K. Townsend, ``{World Sheet supersymmetry and anomaly
  cancellation in the heterotic string},''
{\em Phys. Lett.} {\bf B178} (1986)  187.

\bibitem{Atiyah:1978wi}
M.~Atiyah, N.~J. Hitchin, and I.~Singer, ``{Selfduality in four-dimensional
  Riemannian geometry},''
{\em Proc.Roy.Soc.Lond.} {\bf A362} (1978)  425--461.

\bibitem{Eguchi:1980jx}
T.~Eguchi, P.~B. Gilkey, and A.~J. Hanson, ``{Gravitation, gauge theories and
  differential geometry},''
\href{http://dx.doi.org/10.1016/0370-1573(80)90130-1}{{\em Phys.Rept.} {\bf 66}
  (1980)  213}.

\bibitem{MR0440554}
J.~W. Milnor and J.~D. Stasheff, {\em Characteristic classes}.
\newblock Princeton University Press, Princeton, N. J., 1974.
\newblock Annals of Mathematics Studies, No. 76.

\bibitem{Witten:1985ga}
E.~Witten, ``Global anomalies in string theory,'' in {\em Argonne symposium on
  geometry, anomalies and topology}, W.~A. Bardeen, ed., Argonne.
\newblock 1985.

\bibitem{Freed:1986zx}
D.~Freed, ``Determinants, torsion, and strings,''
\href{http://dx.doi.org/10.1007/BF01221001}{{\em Commun.Math.Phys.} {\bf 107}
  (1986)  483--513}.

\bibitem{Distler:1986wm}
J.~Distler, ``Resurrecting (2,0) compactifications,''
{\em Phys. Lett.} {\bf B188} (1987)  431--436.

\bibitem{Lawson:1989sp}
H.~B. Lawson, Jr. and M.-L. Michelsohn, {\em Spin geometry}, vol.~38 of {\em
  Princeton Mathematical Series}.
\newblock Princeton University Press, Princeton, NJ, 1989.

\bibitem{Melnikov:2010pq}
I.~V. Melnikov and R.~Minasian, ``{Heterotic sigma models with N=2 space-time
  supersymmetry},'' \href{http://dx.doi.org/10.1007/JHEP09(2011)065}{{\em JHEP}
  {\bf 1109} (2011)  065}, \href{http://arxiv.org/abs/1010.5365}{{\tt
  arXiv:1010.5365 [hep-th]}}.

\bibitem{Joyce:2007rh}
D.~D. Joyce, {\em Riemannian holonomy groups and calibrated geometry}, vol.~12
  of {\em Oxford Graduate Texts in Mathematics}.
\newblock Oxford University Press, Oxford, 2007.

\bibitem{Boyer:1988hm}
C.~P. Boyer, ``A note on hyper-{H}ermitian four-manifolds,'' {\em Proc. Amer.
  Math. Soc.} {\bf 102} (1988) no.~1, 157--164.

\bibitem{Becker:2008rc}
M.~Becker, L.-S. Tseng, and S.-T. Yau, ``{New Heterotic Non-Kahler
  Geometries},''
\href{http://arxiv.org/abs/0807.0827}{{\tt arXiv:0807.0827 [hep-th]}}.

\bibitem{Carlevaro:2011mn}
D.~Israel and L.~Carlevaro, ``{Local models of heterotic flux vacua: Spacetime
  and worldsheet aspects},'' {\em Fortsch.Phys.} {\bf 59} (2011)  716--722,
  \href{http://arxiv.org/abs/1109.1534}{{\tt arXiv:1109.1534 [hep-th]}}.

\bibitem{Becker:2006et}
K.~Becker, M.~Becker, J.-X. Fu, L.-S. Tseng, and S.-T. Yau, ``{Anomaly
  cancellation and smooth non-Kaehler solutions in heterotic string theory},''
  {\em Nucl. Phys.} {\bf B751} (2006)  108--128,
\href{http://arxiv.org/abs/hep-th/0604137}{{\tt arXiv:hep-th/0604137}}.

\bibitem{Kodaira:1981cx}
K.~Kodaira, {\em Complex manifolds and deformation of complex structures}.
\newblock Classics in Mathematics. Springer-Verlag, Berlin, 2005.

\bibitem{Strominger:1986uh}
A.~Strominger, ``{Superstrings with torsion},''
{\em Nucl. Phys.} {\bf B274} (1986)  253.

\bibitem{Hull:1986kz}
C.~Hull, ``{Compactifications of the Heterotic Superstring},''
\href{http://dx.doi.org/10.1016/0370-2693(86)91393-6}{{\em Phys.Lett.} {\bf
  B178} (1986)  357}.

\bibitem{Nibbelink:2012wb}
S.~G. Nibbelink and L.~Horstmeyer, ``{Super Weyl invariance: BPS equations from
  heterotic worldsheets},''
\href{http://arxiv.org/abs/1203.6827}{{\tt arXiv:1203.6827 [hep-th]}}.

\bibitem{Dasgupta:1999ss}
K.~Dasgupta, G.~Rajesh, and S.~Sethi, ``{M theory, orientifolds and G-flux},''
  {\em JHEP} {\bf 08} (1999)  023,
\href{http://arxiv.org/abs/hep-th/9908088}{{\tt arXiv:hep-th/9908088}}.

\bibitem{Fu:2006vj}
J.-X. Fu and S.-T. Yau, ``{The theory of superstring with flux on non-Kaehler
  manifolds and the complex Monge-Ampere equation},'' {\em J. Diff. Geom.} {\bf
  78} (2009)  369--428,
\href{http://arxiv.org/abs/hep-th/0604063}{{\tt arXiv:hep-th/0604063}}.

\bibitem{Witten:1999eg}
E.~Witten, ``{World sheet corrections via D instantons},'' {\em JHEP} {\bf
  0002} (2000)  030,
\href{http://arxiv.org/abs/hep-th/9907041}{{\tt arXiv:hep-th/9907041
  [hep-th]}}.

\bibitem{Alvarez:1984es}
O.~Alvarez, ``Topological quantization and cohomology,''
  \href{http://dx.doi.org/10.1007/BF01212452}{{\em Commun.Math.Phys.} {\bf 100}
  (1985)  279}.

\bibitem{Rohm:1985jv}
R.~Rohm and E.~Witten, ``{The antisymmetric tensor field in superstring
  theory},'' \href{http://dx.doi.org/10.1016/0003-4916(86)90099-0}{{\em Annals
  Phys.} {\bf 170} (1986)  454}.

\bibitem{Hopkins:2002rd}
M.~Hopkins and I.~Singer, ``{Quadratic functions in geometry, topology, and M
  theory},'' {\em J.Diff.Geom.} {\bf 70} (2005)  329--452,
  \href{http://arxiv.org/abs/math/0211216}{{\tt arXiv:math/0211216 [math-at]}}.

\bibitem{Bershadsky:1996nh}
M.~Bershadsky, K.~A. Intriligator, S.~Kachru, D.~R. Morrison, V.~Sadov, {\em et
  al.}, ``{Geometric singularities and enhanced gauge symmetries},'' {\em
  Nucl.Phys.} {\bf B481} (1996)  215--252,
  \href{http://arxiv.org/abs/hep-th/9605200}{{\tt arXiv:hep-th/9605200
  [hep-th]}}.

\bibitem{Honecker:2006dt}
G.~Honecker, ``{Massive U(1)s and heterotic five-branes on K3},'' {\em
  Nucl.Phys.} {\bf B748} (2006)  126--148,
  \href{http://arxiv.org/abs/hep-th/0602101}{{\tt arXiv:hep-th/0602101
  [hep-th]}}.

\bibitem{MR658473}
C.~H. Taubes, ``Self-dual {Y}ang-{M}ills connections on non-self-dual
  {$4$}-manifolds,'' {\em J. Differential Geom.} {\bf 17} (1982) no.~1,
  139--170.

\bibitem{Freed:1991in}
D.~S. Freed and K.~K. Uhlenbeck, {\em Instantons and four-manifolds}.
\newblock Mathematical Sciences Research Institute Publications.
  Springer-Verlag, New York, second~ed., 1991.

\bibitem{Donaldson:1990bk}
S.~K. Donaldson and P.~B. Kronheimer, {\em The geometry of four-manifolds}.
\newblock Oxford Mathematical Monographs. The Clarendon Press Oxford University
  Press, New York, 1990.

\bibitem{Witten:1999fq}
E.~Witten, ``{Heterotic string conformal field theory and A-D-E
  singularities},'' {\em JHEP} {\bf 02} (2000)  025,
\href{http://arxiv.org/abs/hep-th/9909229}{{\tt arXiv:hep-th/9909229}}.

\bibitem{MR2172498}
D.~Huybrechts, ``Moduli spaces of hyperk\"ahler manifolds and mirror
  symmetry,'' in {\em Intersection theory and moduli}, ICTP Lect. Notes, XIX,
  pp.~185--247 (electronic).
\newblock Abdus Salam Int. Cent. Theoret. Phys., Trieste, 2004.

\bibitem{Dine:1987xk}
M.~Dine, N.~Seiberg, and E.~Witten, ``{Fayet-Iliopoulos terms in string
  theory},''
{\em Nucl. Phys.} {\bf B289} (1987)  589.

\bibitem{Honecker:2006qz}
G.~Honecker and M.~Trapletti, ``{Merging Heterotic Orbifolds and K3
  Compactifications with Line Bundles},''
  \href{http://dx.doi.org/10.1088/1126-6708/2007/01/051}{{\em JHEP} {\bf 0701}
  (2007)  051},
\href{http://arxiv.org/abs/hep-th/0612030}{{\tt arXiv:hep-th/0612030
  [hep-th]}}.

\bibitem{Donagi:2009ra}
R.~Donagi and M.~Wijnholt, ``{Higgs bundles and UV completion in F-theory},''
  \href{http://arxiv.org/abs/0904.1218}{{\tt arXiv:0904.1218 [hep-th]}}.

\bibitem{Anderson:2011ty}
L.~B. Anderson, J.~Gray, A.~Lukas, and B.~Ovrut, ``{The Atiyah class and
  complex structure stabilization in heterotic Calabi-Yau compactifications},''
  \href{http://arxiv.org/abs/1107.5076}{{\tt arXiv:1107.5076 [hep-th]}}.

\bibitem{Melnikov:2011ez}
I.~V. Melnikov and E.~Sharpe, ``{On marginal deformations of (0,2) non-linear
  sigma models},'' \href{http://dx.doi.org/10.1016/j.physletb.2011.10.055}{{\em
  Phys.Lett.} {\bf B705} (2011)  529--534},
\href{http://arxiv.org/abs/1110.1886}{{\tt arXiv:1110.1886 [hep-th]}}.

\bibitem{Distler:1987ee}
J.~Distler and B.~R. Greene, ``{Aspects of (2,0) string compactifications},''
{\em Nucl. Phys.} {\bf B304} (1988)  1.

\bibitem{Griffiths:1978pa}
P.~Griffiths and J.~Harris, {\em Principles of algebraic geometry}.
\newblock Wiley-Interscience [John Wiley \& Sons], New York, 1978.
\newblock Pure and Applied Mathematics.

\bibitem{Barth:2004ne}
W.~P. Barth, K.~Hulek, C.~A.~M. Peters, and A.~Van~de Ven, {\em Compact complex
  surfaces}, vol.~4.
\newblock Springer-Verlag, Berlin, second~ed., 2004.

\bibitem{Kumar:2009zc}
V.~Kumar and W.~Taylor, ``{Freedom and Constraints in the K3 Landscape},''
  \href{http://dx.doi.org/10.1088/1126-6708/2009/05/066}{{\em JHEP} {\bf 0905}
  (2009)  066},
\href{http://arxiv.org/abs/0903.0386}{{\tt arXiv:0903.0386 [hep-th]}}.

\bibitem{MR1228584}
K.~Oguiso, ``On algebraic fiber space structures on a {C}alabi-{Y}au
  {$3$}-fold,'' {\em Internat. J. Math.} {\bf 4} (1993) no.~3, 439--465. With
  an appendix by Noboru Nakayama.

\bibitem{Morrison:1996na}
D.~R. Morrison and C.~Vafa, ``{Compactifications of F theory on Calabi-Yau
  threefolds. 1},'' \href{http://dx.doi.org/10.1016/0550-3213(96)00242-8}{{\em
  Nucl.Phys.} {\bf B473} (1996)  74--92},
  \href{http://arxiv.org/abs/hep-th/9602114}{{\tt arXiv:hep-th/9602114
  [hep-th]}}.

\bibitem{Ferrara:1996wv}
S.~Ferrara, R.~Minasian, and A.~Sagnotti, ``{Low-energy analysis of M and F
  theories on Calabi-Yau threefolds},'' {\em Nucl.Phys.} {\bf B474} (1996)
  323--342, \href{http://arxiv.org/abs/hep-th/9604097}{{\tt
  arXiv:hep-th/9604097 [hep-th]}}.

\bibitem{Avram:1996pj}
A.~Avram, M.~Kreuzer, M.~Mandelberg, and H.~Skarke, ``{Searching for K3
  fibrations},'' \href{http://dx.doi.org/10.1016/S0550-3213(97)00214-9}{{\em
  Nucl.Phys.} {\bf B494} (1997)  567--589},
  \href{http://arxiv.org/abs/hep-th/9610154}{{\tt arXiv:hep-th/9610154
  [hep-th]}}.

\bibitem{Klemm:2004km}
A.~Klemm, M.~Kreuzer, E.~Riegler, and E.~Scheidegger, ``{Topological string
  amplitudes, complete intersection Calabi-Yau spaces and threshold
  corrections},'' \href{http://dx.doi.org/10.1088/1126-6708/2005/05/023}{{\em
  JHEP} {\bf 0505} (2005)  023},
  \href{http://arxiv.org/abs/hep-th/0410018}{{\tt arXiv:hep-th/0410018
  [hep-th]}}.

\bibitem{Kaplunovsky:1995tm}
V.~Kaplunovsky, J.~Louis, and S.~Theisen, ``{Aspects of duality in N=2 string
  vacua},'' \href{http://dx.doi.org/10.1016/0370-2693(95)00857-H}{{\em
  Phys.Lett.} {\bf B357} (1995)  71--75},
  \href{http://arxiv.org/abs/hep-th/9506110}{{\tt arXiv:hep-th/9506110
  [hep-th]}}.

\bibitem{Candelas:1993dm}
P.~Candelas, X.~De~La~Ossa, A.~Font, S.~H. Katz, and D.~R. Morrison, ``{Mirror
  symmetry for two parameter models. I},'' {\em Nucl. Phys.} {\bf B416} (1994)
  481--538,
\href{http://arxiv.org/abs/hep-th/9308083}{{\tt arXiv:hep-th/9308083}}.

\bibitem{Hosono:1993qy}
S.~Hosono, A.~Klemm, S.~Theisen, and S.-T. Yau, ``{Mirror symmetry, mirror map
  and applications to Calabi-Yau hypersurfaces},'' {\em Commun. Math. Phys.}
  {\bf 167} (1995)  301--350,
\href{http://arxiv.org/abs/hep-th/9308122}{{\tt arXiv:hep-th/9308122}}.

\bibitem{Kreuzer:2002uu}
M.~Kreuzer and H.~Skarke, ``{PALP: A Package for analyzing lattice polytopes
  with applications to toric geometry},'' {\em Comput. Phys. Commun.} {\bf 157}
  (2004)  87--106,
\href{http://arxiv.org/abs/math/0204356}{{\tt arXiv:math/0204356}}.

\bibitem{Aldazabal:1995yw}
G.~Aldazabal, A.~Font, L.~E. Ibanez, and F.~Quevedo, ``{Chains of N=2, D = 4
  heterotic type II duals},'' {\em Nucl.Phys.} {\bf B461} (1996)  85--100,
  \href{http://arxiv.org/abs/hep-th/9510093}{{\tt arXiv:hep-th/9510093}}.

\bibitem{Candelas:1996su}
P.~Candelas and A.~Font, ``{Duality between the webs of heterotic and type II
  vacua},'' {\em Nucl.Phys.} {\bf B511} (1998)  295--325,
  \href{http://arxiv.org/abs/hep-th/9603170}{{\tt arXiv:hep-th/9603170
  [hep-th]}}.

\bibitem{Duff:1996rs}
M.~Duff, R.~Minasian, and E.~Witten, ``{Evidence for heterotic / heterotic
  duality},'' \href{http://dx.doi.org/10.1016/0550-3213(96)00059-4}{{\em
  Nucl.Phys.} {\bf B465} (1996)  413--438},
\href{http://arxiv.org/abs/hep-th/9601036}{{\tt arXiv:hep-th/9601036
  [hep-th]}}.

\bibitem{Evslin:2008zm}
J.~Evslin and R.~Minasian, ``{Topology change from (heterotic) Narain
  T-duality},'' {\em Nucl. Phys.} {\bf B820} (2009)  213--236,
\href{http://arxiv.org/abs/0811.3866}{{\tt arXiv:0811.3866 [hep-th]}}.

\bibitem{Gawedzki:1988hq}
K.~Gawedzki and A.~Kupiainen, ``{G/H} conformal field theory from gauged {WZW}
  model,'' {\em Phys.Lett.} {\bf B215} (1988)  119--123.

\bibitem{Witten:1991mm}
E.~Witten, ``{On Holomorphic factorization of WZW and coset models},''
\href{http://dx.doi.org/10.1007/BF02099196}{{\em Commun.Math.Phys.} {\bf 144}
  (1992)  189--212}.

\bibitem{Berglund:1995dv}
P.~Berglund, C.~V. Johnson, S.~Kachru, and P.~Zaugg, ``{Heterotic coset models
  and (0,2) string vacua},''
  \href{http://dx.doi.org/10.1016/0550-3213(95)00641-9}{{\em Nucl.Phys.} {\bf
  B460} (1996)  252--298}, \href{http://arxiv.org/abs/hep-th/9509170}{{\tt
  arXiv:hep-th/9509170 [hep-th]}}.

\bibitem{Adams:2009av}
A.~Adams and D.~Guarrera, ``{Heterotic Flux Vacua from Hybrid Linear Models},''
\href{http://arxiv.org/abs/0902.4440}{{\tt arXiv:0902.4440 [hep-th]}}.

\bibitem{Rohm:1984ix}
R.~Rohm, ``Anomalous interactions for the supersymmetric nonlinear sigma model
  in two-dimensions,''
\href{http://dx.doi.org/10.1103/PhysRevD.32.2849}{{\em Phys.Rev.} {\bf D32}
  (1985)  2849}.

\bibitem{Witten:1991mk}
E.~Witten, ``{The N matrix model and gauged WZW models},''
{\em Nucl. Phys.} {\bf B371} (1992)  191--245.

\bibitem{Salamon:2002hg}
S.~M. Salamon, ``Hermitian geometry,'' in {\em Invitations to geometry and
  topology}, vol.~7 of {\em Oxf. Grad. Texts Math.}, pp.~233--291.
\newblock Oxford Univ. Press, Oxford, 2002.

\bibitem{McOrist:2010jw}
J.~McOrist, D.~R. Morrison, and S.~Sethi, ``{Geometries, Non-Geometries, and
  Fluxes},''
\href{http://arxiv.org/abs/1004.5447}{{\tt arXiv:1004.5447 [hep-th]}}.

\bibitem{Sethi:1996es}
S.~Sethi, C.~Vafa, and E.~Witten, ``{Constraints on low dimensional string
  compactifications},''
  \href{http://dx.doi.org/10.1016/S0550-3213(96)00483-X}{{\em Nucl.Phys.} {\bf
  B480} (1996)  213--224},
\href{http://arxiv.org/abs/hep-th/9606122}{{\tt arXiv:hep-th/9606122
  [hep-th]}}.

\bibitem{Aspinwall:2005qw}
P.~S. Aspinwall, ``{An analysis of fluxes by duality},''
\href{http://arxiv.org/abs/hep-th/0504036}{{\tt arXiv:hep-th/0504036
  [hep-th]}}.

\bibitem{Ketov:2000qn}
S.~Ketov, {\em {Quantum non-linear sigma models}}.
\newblock Springer, 2000.

\bibitem{Green:1987qu}
M.~B. Green and N.~Seiberg, ``Contact interactions in superstring theory,''
\href{http://dx.doi.org/10.1016/0550-3213(88)90549-4}{{\em Nucl.Phys.} {\bf
  B299} (1988)  559}.

\bibitem{Friedan:1980jm}
D.~H. Friedan, ``{Nonlinear models in two + epsilon dimensions},''
  \href{http://dx.doi.org/10.1016/0003-4916(85)90384-7}{{\em Annals Phys.} {\bf
  163} (1985)  318}.
Ph.D. Thesis.

\bibitem{AlvarezGaume:1981hn}
L.~Alvarez-Gaume, D.~Z. Freedman, and S.~Mukhi, ``{The Background field method
  and the ultraviolet structure of the supersymmetric nonlinear sigma model},''
\href{http://dx.doi.org/10.1016/0003-4916(81)90006-3}{{\em Annals Phys.} {\bf
  134} (1981)  85}.

\bibitem{Seiberg:1994rs}
N.~Seiberg and E.~Witten, ``{Electric - magnetic duality, monopole
  condensation, and confinement in N=2 supersymmetric Yang-Mills theory},''
  {\em Nucl. Phys.} {\bf B426} (1994)  19--52,
\href{http://arxiv.org/abs/hep-th/9407087}{{\tt arXiv:hep-th/9407087}}.

\bibitem{Seiberg:1994aj}
N.~Seiberg and E.~Witten, ``{Monopoles, duality and chiral symmetry breaking in
  N=2 supersymmetric QCD},'' {\em Nucl. Phys.} {\bf B431} (1994)  484--550,
\href{http://arxiv.org/abs/hep-th/9408099}{{\tt arXiv:hep-th/9408099}}.

\bibitem{Argyres:1996eh}
P.~C. Argyres, M.~R. Plesser, and N.~Seiberg, ``{The Moduli space of vacua of
  N=2 SUSY QCD and duality in N=1 SUSY QCD},'' {\em Nucl.Phys.} {\bf B471}
  (1996)  159--194, \href{http://arxiv.org/abs/hep-th/9603042}{{\tt
  arXiv:hep-th/9603042 [hep-th]}}.

\bibitem{Argyres:1996hc}
P.~C. Argyres, M.~R. Plesser, and A.~D. Shapere, ``{N=2 moduli spaces and N=1
  dualities for SO(n(c)) and USp(2n(c)) superQCD},''
  \href{http://dx.doi.org/10.1016/S0550-3213(96)00583-4}{{\em Nucl.Phys.} {\bf
  B483} (1997)  172--186}, \href{http://arxiv.org/abs/hep-th/9608129}{{\tt
  arXiv:hep-th/9608129 [hep-th]}}.

\bibitem{Benini:2009gi}
F.~Benini, S.~Benvenuti, and Y.~Tachikawa, ``{Webs of five-branes and N=2
  superconformal field theories},'' {\em JHEP} {\bf 0909} (2009)  052,
  \href{http://arxiv.org/abs/0906.0359}{{\tt arXiv:0906.0359 [hep-th]}}.

\bibitem{Hanany:2010qu}
A.~Hanany and N.~Mekareeya, ``{Tri-vertices and SU(2)'s},''
  \href{http://dx.doi.org/10.1007/JHEP02(2011)069}{{\em JHEP} {\bf 1102} (2011)
   069}, \href{http://arxiv.org/abs/1012.2119}{{\tt arXiv:1012.2119 [hep-th]}}.

\bibitem{Chacaltana:2012zy}
O.~Chacaltana, J.~Distler, and Y.~Tachikawa, ``{Nilpotent orbits and
  codimension-two defects of 6d N=(2,0) theories},''
  \href{http://arxiv.org/abs/1203.2930}{{\tt arXiv:1203.2930 [hep-th]}}.

\bibitem{Kugo:1994qr}
T.~Kugo and J.~Sato, ``{Dynamical symmetry breaking in an {$E_6$} GUT model},''
  {\em Prog.Theor.Phys.} {\bf 91} (1994)  1217--1238,
  \href{http://arxiv.org/abs/hep-ph/9402357}{{\tt arXiv:hep-ph/9402357}}.

\bibitem{Kephart:1981gf}
T.~W. Kephart and M.~T. Vaughn, ``Tensor methods for the exceptional group
  {$E_6$},'' {\em Annals Phys.} {\bf 145} (1983)  162.

\bibitem{Gursey:1980}
F.~Gursey, ``Symmetry breaking patterns in {$E_6$}.'' Invited talk at the april
  1980 new hampshire workshop on grand unified theories.

\end{thebibliography}
\providecommand{\href}[2]{#2}\begingroup\raggedright\endgroup

\end{document}